\documentclass[aps,prd,twocolumn,groupedaddress,showpacs]{revtex4}

\usepackage{mathrsfs}
\usepackage{bm}
\usepackage{stmaryrd}
\usepackage{slashed}
\usepackage{amsfonts}

\begin{document}


\title{Liouville equations for neutrino distribution matrices}



\author{Christian Y. Cardall}
\email[]{cardallcy@ornl.gov}
\affiliation{Physics Division, Oak Ridge National Laboratory, Oak Ridge,
 	TN 37831-6354}
\affiliation{Department of Physics and Astronomy, University of Tennessee,
	Knoxville, TN 37996-1200} 


\date{\today}

\begin{abstract}
The classical notion of a single-particle scalar distribution function or phase space density can be generalized to a matrix in order to accommodate superpositions of states of discrete quantum numbers, such as neutrino mass/flavor. Such a `neutrino distribution matrix' is thus an appropriate construct to describe a neutrino gas that may vary in space as well as time and in which flavor mixing competes with collisions. The Liouville equations obeyed by relativistic neutrino distribution matrices, including the spatial derivative and vacuum flavor mixing terms, can be explicitly but elegantly derived in two new ways: from a covariant version of the familiar simple model of flavor mixing, and from the Klein-Gordon equations satisfied by a quantum `density function' (mean value of paired quantum field operators). Associated with the latter derivation is a case study in how the joint position/momentum dependence of a classical gas (albeit with Fermi statistics) emerges from a formalism built on quantum fields.
\end{abstract}


\maketitle

\def\neutrinoKet#1#2{ \left|\nu_{\mathscr W} #1;#2 \right\rangle }
\def\antineutrinoKet#1#2{ \left|\bar\nu_{\mathscr W} #1;#2 \right\rangle }

\def\neutrinoBra#1#2{ \left\langle\nu_{\mathscr W} #1;#2 \right| }
\def\antineutrinoBra#1#2{ \left\langle\bar\nu_{\mathscr W} #1;#2 \right| }

\def\neutrinoHalfBra#1#2{ \left\langle\nu_{\mathscr W} #1;#2 \right. }
\def\antineutrinoHalfBra#1#2{ \left\langle\bar\nu_{\mathscr W} #1;#2 \right. }

\section{Introduction}
\label{sec:introduction}

The decoupling of neutrinos from dense nuclear matter occurs in a number of environments in which neutrino flavor mixing may also play an important role, including the early universe \cite{Prakash2001Neutrino-Propag,Dolgov2002Neutrinos-in-co} and core-collapse supernovae \cite{Fuller2006Simultaneous-fl,Duan2006Collective-neut,Duan2006Simulation-of-c}. Treatment of the neutrinos' transition from diffusion to free streaming requires some sort of `transport calculation' embodying the principles of some kind of `kinetic theory.' 

Localized `microscopic' quantum mechanical effects can be handled within the framework of classical kinetic theory \cite{Lindquist1966Relativistic-Tr,Ehlers1971General-Relativ,Israel1972The-Relativisti}. In classical kinetic theory a single-particle distribution function $f(t,{\bf x},{\bf p})$ quantifies the average number $dN$ of particles of a particular type, having spin degeneracy $g$ and momenta within $d^3{\bf p}$ of ${\bf p}$, at positions within $d^3{\bf x}$ of ${\bf x}$:
\begin{equation}
dN = f(t,{\bf x},{\bf p})\; \frac{g\, d^3{\bf p}}{(2\pi)^3} \; d^3{\bf x}. \label{classicalDistribution}
\end{equation}
The Boltzmann equation equates a collision integral $C(f)$ to the rate of change $df/d\lambda$ of the average density of particles in classical phase space trajectories $\left(x(\lambda),{\bf p}(\lambda)\right)$. 
Given the geodesic equations defining worldlines $x(\lambda)$, 
\begin{eqnarray}
\frac{dx^\mu}{d\lambda} &=& p^\mu, \\
\frac{dp^\mu}{d\lambda} &=& -{\Gamma^\mu}_{\nu\rho}\,p^\nu p^\rho
\end{eqnarray}
(here ${\Gamma^\mu}_{\nu\rho}$ are the connection coefficients associated with the spacetime metric), the Boltzmann equation can be expressed
\begin{equation}
p^\mu \frac{\partial f}{\partial x^\mu} -{\Gamma^i}_{\nu\rho}\,p^\nu p^\rho\,\frac{\partial f}{\partial p^{i}}= C(f). \label{boltzmann}
\end{equation}
On the left-hand side the Liouville operator acts upon $f$. The collision integral on the right-hand side represents the phase space density (rate per space volume per momentum space volume) of isolated  `microscopic' or point-like scattering events between classical trajectories. By reasonable extension, the collision integral also includes inherently quantum-mechanical processes affecting the population of classical phase space trajectories, such as particle decays, particle emission/absorption, and pair creation/annihilation. The restriction to isolated point-like transitions allows for insertion into $C(f)$ of interaction rates computed (for instance) with the standard methods of quantum field theory, together with factors $1\pm f$ (with upper sign for bosons and lower sign for fermions) encoding the impact of quantum statistics upon available final-state phase space.

However, neutrino flavor mixing is a `macroscopic' quantum mechanical effect, requiring the evolution of amplitudes across time and/or distance scales comparable to scales characteristic of the total system under consideration; hence flavor mixing cannot in general be described by the scalar distribution function $f(t,{\bf x},{\bf p})$ and the Boltzmann equation it obeys, these being concerned only with the evolution of particle number, and not quantum amplitudes and the evolution of phases.

Classical transport being inadequate as a conceptual framework for handling flavor mixing, attention turns to a statistical treatment of a quantum gas. An approach especially suited to cases of spatial homogeneity begins with the definition of a quantum occupation number $n(t,{\bf q})$, which quantifies the average number $dN$ of particles of spin degeneracy $g$, occupying momentum eigenstates within $d^3{\bf q}$ of ${\bf q}$ within the (effectively infinite) quantization volume $V$:  
\begin{equation}
dN = n(t, {\bf q})\;  \frac{g\, V d^3{\bf q}}{(2\pi)^3}. \label{occupationNumber}
\end{equation}
The time derivative $dn/dt$ is equal to an entity similar to the collision integral in the Boltzmann equation, constructed using transition rates between quantum states.

Distinctions between quantum occupation numbers $n(t,{\bf q})$ and classical distribution functions $f(t,{\bf x},{\bf p})$ should be kept in mind. Because they specify the average populations of quantum states rather than classical trajectories, quantum occupation numbers are distinguished from classical distribution functions  by an absence of spatial dependence, as required by the impossibility in quantum mechanics of simultaneous sharp specification of both position and momentum. Moreover, there is a subtle difference between the momenta ${\bf p}$ in $f(t,{\bf x},{\bf p})$ and ${\bf q}$ in $n(t,{\bf q})$. In the case of an occupation number, ${\bf q}$ is the eigenvalue of a momentum eigenstate. But in a classical distribution function, ${\bf p}$ represents the momentum of a classical particle; considered as a limit of a quantum mechanical description, ${\bf p}$ might therefore be thought of as representing the centroid of a momentum-space wave packet---that is, ${\bf p}$ is the expectation value of a superposition of momentum eigenstates ${\bf q}$.  

Accommodation of flavor mixing in a neutrino gas was first contemplated \cite{Dolgov1981Neutrinos-in-th,Barbieri1991Neutrino-Oscill} for the (homogeneous and isotropic) early universe, through the introduction of what might at first glance be thought of as a `neutrino (quantum) occupation matrix' $\rho(t,|{\bf q}|)$. This is ``a matrix in the space of neutrino species'' \cite{Dolgov1981Neutrinos-in-th}, whose diagonal elements are occupation numbers of the various neutrino species, and whose off-diagonal elements quantify the extent to which neutrinos exist in superpositions of distinct species. Its introduction appears to have been motivated by the recognition that neutrino interaction amplitudes constitute a matrix with nonvanishing off-diagonal entries when written in terms of `physical states' (neutrinos of definite mass). The fact that the off-diagonal elements of a neutrino occupation matrix $\rho(t,|{\bf q}|)$ contain information on coherent superpositions is reminiscent of a density matrix, as is the fact that it obeys a Heisenberg-like equation of motion (time derivative given by a commutator with a Hamiltonian). Indeed it has sometimes been called a `density matrix' in the literature (for instance, in Refs. \cite{Dolgov1981Neutrinos-in-th,Barbieri1991Neutrino-Oscill,Qian1995Neutrino-neutri}). However, it is perhaps better thought of as a `matrix of densities' \cite{Sigl1993General-kinetic}, since it is not a density matrix (as traditionally defined) for multiparticle neutrino 
states, in that its
trace is not equal to unity. Hence its diagonal elements are not `probabilities' in the strictest (that is, absolute) sense \footnote{On the other hand, the diagonal elements of $\rho(t,|{\bf q}|)$ might be considered `probabilities' in the same sense that (for instance) the Maxwell-Boltzmann distribution is a `probability distribution,' that is, one that is normalized to a total number of particles rather than unity. Moreover, there would seem to be a conceptual and historical linkage---suggested by the shared term `density'---between the notions of a phase space density and a density matrix. Indeed the classical and quantum situations are similar, in that by neglecting correlations one moves from {\em absolute} probabilities in a multiparticle phase space (classical case) or Hilbert space (quantum case) to {\em relative} probabilities normalized to a total number of particles, through the introduction of a single-particle phase space density (classical case) or occupation number (quantum case).}\footnote{A variant formalism developed in Ref. \cite{McKellar1994Oscillating-neu} does use a density matrix as traditionally defined, that is, one whose trace is equal to unity. This is accomplished in a context of neutrino emission/absorption and pair creation/annihilation by including charged leptons in the degrees of freedom spanned by a density matrix for a single particle. Still excluding multiparticle correlations by fiat, and restricting attention to systems sufficiently dilute that the effects of quantum statistics can be ignored, the density operator for the entire collection of $N$ leptons+antileptons is taken to be a tensor product of $N$ single-particle density operators.}\footnote{In the case of $\rho(t,|{\bf q}|)$ some authors choose to use the term `density matrix' together with a modifier. In these cases the object that corresponds to $\rho(t,|{\bf q}|)$ is taken to be ${\rm Tr}(\hat \rho_{\rm tot} \, a_{\beta,{\bf q}}^\dagger a_{\alpha,{\bf q}})$. This is an expectation value of what looks like a number operator---that is, a product of a creation operator and an annihilation operator in Fock space---but with these creation and annihilation operators representing (perhaps distinct) species $\alpha$ and $\beta$. The expectation value is taken with respect to the `complete' density operator (density matrix) $\hat \rho_{\rm tot}$ for the total system, encompassing the entire multiparticle Fock space of all particle types. Works in which an object like ${\rm Tr}(\hat \rho_{\rm tot} \, a_{\beta,{\bf q}}^\dagger a_{\alpha,{\bf q}})$ is written down to represent neutrino ensembles include Ref. \cite{Rudzsky1990Kinetic-Equatio}, where it is called the `one-particle density matrix in momentum representation'; and Ref. \cite{Prakash2001Neutrino-Propag}, where it is called the `momentum-flavor density matrix.' }.

In typical studies of the epoch of big-bang nucleosynthesis, however, the momenta apparently take on their classical significance---that is, there seems to be a (generally tacit) assumption that the neutrino gas is described by a `distribution matrix' $\rho(t,|{\bf p}|)$ rather than an `occupation matrix' $\rho(t,|{\bf q}|)$. This is because the expansion of the universe must be accounted for. Classical kinetic theory calculations in the context of the early universe---which do not involve flavor mixing---have long used the Boltzmann equation, or a momentum integral thereof. The redshift (or number dilution, in the momentum-integrated case) due to cosmological expansion results from nonvanishing connection coefficients in Eq. (\ref{boltzmann}); see for instance Ref. \cite{Kolb1990The-Early-Unive}. The very reasonable, if unremarked \cite{Dolgov1981Neutrinos-in-th,Barbieri1991Neutrino-Oscill}, assumption  seems to be that a replacement
\begin{equation}
\frac{d\rho(t,|{\bf q}|)}{dt} \rightarrow \frac{\partial\rho(t,|{\bf p}|)}{\partial t}- H |{\bf p}| \frac{\partial\rho(t,|{\bf p}|)}{\partial |{\bf p}|} \label{replacement}
\end{equation}
obtains, where the Hubble parameter $H$ is the cosmological expansion rate. That is,
the total time derivative $d\rho/dt$ of a quantum occupation matrix $\rho(t,|{\bf q}|)$ that satisfies a Heisenberg-like equation of motion somehow goes over to the action of the classical Liouville operator upon a distribution matrix $\rho(t,|{\bf p}|)$ that is classical in all but the discrete quantum numbers (e.g. flavor/mass) responsible for the matrix structure. The intuition behind this replacement is evidently similar to that which motivates the use of interaction rates computed with quantum field theory (plus quantum statistics) in the Boltzmann equation's collision integral, as mentioned above in connection with classical kinetic theory. In particular, the `Liouville replacement' of Eq. (\ref{replacement}) and the calculation of interaction rates that go into collision integrals share the following feature: plane waves (momentum eigenstates), here labeled by ${\bf q}$, are taken as proxies for classical particles (or quantum wave packets) with momenta (or momentum-space wave packet centroids) ${\bf p}$. 

The subtle distinction between `quantum' momenta ${\bf q}$ and `classical' momenta ${\bf p}$ is hardly noticeable and easily glossed over in the treatment of a spatially homogeneous system like the early universe, but it becomes more obvious upon consideration of spatial dependence. This is because writing down an expression like $\rho(t,{\bf x},{\bf q})$ immediately brings to mind the quantum mechanical incompatibility of position and momentum.  In a limit in which the neutrinos' motion through spacetime is expected to be classical, we want instead an object $\rho(t,{\bf x},{\bf p})$ in which ${\bf p}$ represents something other than quantum numbers of a momentum eigenstate. If approached from a fully quantum perspective, obtaining $\rho(t,{\bf x},{\bf p})$ will be expected to somehow involve a Wigner transformation \cite{Wigner1932On-the-Quantum-} (see also Ref. \cite{Groot1980Relativistic-Ki} for a general treatment, having a somewhat different flavor than that presented in Sec. \ref{sec:correlation} below, that employs a Wigner transformation in connection with relativistic quantum fields). A Wigner transformation is a Fourier transformation with respect to a spatial difference variable, with ${\bf p}$ entering as this difference variable's `Fourier conjugate.' It is also natural (and correct) to guess that a spatial derivative Liouville term $p^i\,\partial \rho / \partial x^i$ would appear in the equation of motion for $\rho(t,{\bf x},{\bf p})$. 

A few previous efforts towards a kinetic theory of neutrinos with flavor mixing have noted the existence of a spatial derivative term in the Liouville operator. These might be divided into two broad classes. In several cases an explicit derivation of this term is absent \cite{Prakash2001Neutrino-Propag,Sigl1993General-kinetic,Rudzsky1990Kinetic-Equatio,Strack2005Generalized-Bol}, but its expected presence in a Heisenberg-like equation of motion is noted, based on appeals to literature in non-relativistic statistical physics \cite{Akhiezer1981Methods-of-Stat,Kampen1992Stochastic-Proc} or quantum optics \cite{Walls1994Quantum-Optics}. The other class includes two works \cite{Sirera1999Relativistic-Wi,Yamada2000Boltzmann-equat} in which spatial derivatives appear in the course of the derivation, thanks to use in one way or another of the Dirac equation obeyed by neutrino quantum field operators. At least from the perspective of working astrophysicists, this second class of approaches has not lent itself to the greatest transparency.

The purpose of this paper is to elucidate the phrase ``somehow goes over to'' in the sentence that follows Eq. (\ref{replacement}), with an emphasis on obtaining the spatial derivative and vacuum flavor mixing terms in the flat spacetime Liouville equations obeyed by relativistic neutrino and antineutrino single-particle distribution matrices (here `relativistic' means that only terms of ${\cal O}(m_\nu^2/ E_\nu)$ are kept, where $m_\nu$ and $E_\nu$ are characteristic neutrino mass and energy scales). Goals include increased simplicity and transparency in comparison with available explicit derivations \cite{Sirera1999Relativistic-Wi,Yamada2000Boltzmann-equat} and a more detailed discussion of the physical interpretation of the Wigner-transformed `density function' (mean value of paired quantum field operators). The value in this lies in an improved understanding of how classical expressions emerge from quantum formalisms. This bridge between the classical and quantum worlds will facilitate study of the potential impact of flavor mixing on the decoupling of neutrinos in (for instance) core-collapse supernovae: the macroscopic quantum effect of flavor mixing must be retained, but the neutrinos' motion through spacetime goes over to a classical description in order that the system's formal dependence on time, position, and momentum be simplified to a degree comparable to that exhibited by a classical single-particle distribution function. (While the formal dependence of the neutrino distributions on time, position, and momentum are not unlike those of a classical single-particle distribution function, there are indications that flavor mixing may lead to new and complicated behavior in the supernova environment  \cite{Fuller2006Simultaneous-fl,Duan2006Collective-neut,Duan2006Simulation-of-c}.)

Two different accounts of the construction of neutrino and antineutrino distribution matrices $\rho(t,{\bf x},{\bf p})$ and $\bar\rho(t,{\bf x},{\bf p})$ and the Liouville equations they obey in the absence of interactions are presented in the following sections. In Sec. \ref{sec:simple} a simple model of the flavor evolution of a single neutrino is taken as a starting point. In this approach the quantum treatment is restricted to flavor evolution; neutrinos are assumed from the outset to follow classical trajectories in spacetime. Distribution matrices are built up from single-neutrino states, and the Liouville equation follows from the evolution equation for these individual states. In contrast, a `density function' constructed from quantum fields, and the equations of motion it obeys, are the basis of the approach presented in Sec. \ref{sec:correlation}. In this case classical expressions involving spacetime variables are not immediately obvious, but can be drawn out through use of a Wigner transformation. Section \ref{sec:conclusion} contains a summary and some remarks looking ahead towards the derivation of interactions from this second formalism. Metric signature $+---$ and units in which $\hbar = c = 1$ are employed throughout.

\section{Starting from a simple model of flavor mixing}
\label{sec:simple}

A simple model of neutrino flavor mixing postulates the existence of flavor and mass eigenstates.
Neutrino and antineutrino flavor eigenstates (labeled by $\alpha$) are related to mass eigenstates (labeled by $i$) by
\begin{eqnarray}
\neutrinoKet{}{\alpha} &=& \sum_i U_{\alpha i}^* \, \neutrinoKet{}{i}, \label{neutrinoBases} \\
\antineutrinoKet{}{\alpha} &=& \sum_i U_{\alpha i} \, \antineutrinoKet{}{i} \label{antineutrinoBases}
\end{eqnarray}
respectively, where $\mathscr{W}$ is the neutrino or antineutrino's classical worldline \footnote{\label{fock}Note that these states are not members of a Fock space, for here momentum is not a quantum number, but has only its classical meaning as the tangent vector to the worldline $\mathscr{W}$. Indeed, the definition of flavor states with momentum promoted to a quantum number is problematic because Fock space creation and annihilation operators obeying the appropriate anticommutation relations cannot be found for nonvanishing mass \cite{Giunti1992Remarks-on-the-}. However, the entire neutrino production/propagation/detection process can be analyzed without invoking the existence of Fock space flavor states. In the relativistic limit and with the satisfaction of other conditions very often realized in practice, flavor change probabilities can be extracted that agree with the results of the simple model. See for instance Ref. \cite{Cardall2000Coherence-of-ne} and references therein.}. These basis transformations feature the same unitary matrix that relates neutrino flavor and mass quantum field operators: $\nu_\alpha(x) = \sum_i U_{\alpha i}\, \nu_i(x)$. 

Consider a `Schr\"odinger picture' in which a neutrino state evolves along a worldline $\mathscr{W}$ with affine parameter $\lambda$. Let $\neutrinoKet{(\lambda)}{\alpha}$ denote a neutrino that was born in flavor $\alpha$ at $\lambda = 0$ and then translated to $\lambda$ by a unitary `worldline evolution operator' $\hat\mathscr{U}_\mathscr{W}(\lambda,0)$. As usual for unitary transformations parametrized by a continuous variable, worldline translations are generated by a Hermitian operator, here denoted $\hat\Lambda_\mathscr{W}$:
\begin{equation}
\hat\mathscr{U}_\mathscr{W}(\lambda+d\lambda, \lambda) = 1 - {\rm i}\,\hat\Lambda_\mathscr{W}\,d\lambda,
\end{equation}
whence the `Schr\"odinger equation'
\begin{equation}
{\rm i} \frac{d}{d\lambda}\neutrinoKet{(\lambda)}{\alpha} = \hat\Lambda_\mathscr{W}\neutrinoKet{(\lambda)}{\alpha}. \label{schrodinger0}
\end{equation}
A definition of $\hat\Lambda_\mathscr{W}$ is needed. In flat spacetime
\begin{equation}
{\rm i} \frac{d}{d\lambda} = {\rm i}\,p_\mathscr{W}^\mu \frac{ \partial}{\partial x^\mu}, \label{ddlambda}
\end{equation}
where the classical four-momentum $p_\mathscr{W}$ is tangent to $\mathscr{W}$. 
The familiar significance of ${\rm i} \, \partial/\partial x^\mu$ as a representation of the generator of spacetime translations motivates the construction of a flavor evolution operator $\hat P_\mathscr{W}^\mu$ modeled on the four-momentum of a particle approaching the relativistic limit:
\begin{eqnarray}
\hat P_\mathscr{W}^0 &=&  \sqrt{ |{\bf p}_\mathscr{W}|^2 + \hat M^2 } \rightarrow |{\bf p_\mathscr{W}}| + \frac{\hat M^2}{2|{\bf p_\mathscr{W}}|}, \\
\hat P_\mathscr{W}^i &=& p_\mathscr{W}^i = {\bf p}_\mathscr{W}^i,
\end{eqnarray}
where the mass operator $\hat M$ with eigenvalues $m_i$ is not diagonal in the flavor basis. Then $\hat\Lambda_\mathscr{W} = p_\mathscr{W}^\mu \hat P_{\mathscr{W}\mu}$, and Eq. (\ref{schrodinger0}) becomes \cite{Cardall1997Neutrino-oscill}
\begin{equation}
{\rm i} \frac{d}{d\lambda}\neutrinoKet{(\lambda)}{\alpha} = \frac{\hat M^2}{2}\neutrinoKet{(\lambda)}{\alpha}, \label{schrodinger1}
\end{equation}
a covariant version of the familiar \cite{Yao2006Review-of-Parti} neutrino flavor evolution equation. Antineutrino states obey the same equation.

The off-diagonal terms in the flavor-basis representation of $\hat M^2$ imply that a Schr\"odinger-picture neutrino state $\neutrinoKet{(\lambda)}{\alpha}$ that begins life in flavor $\alpha$ evolves into a superposition of all flavors $\beta$. The `oscillation probability' $\left| \neutrinoHalfBra{}{\beta} \neutrinoKet{(\lambda)}{\alpha} \right|^2$ for a neutrino flavor transformation $\nu_\alpha \rightarrow \nu_\beta$ that follows from solution of Eq. (\ref{schrodinger1}) is a function of $\lambda$; but it agrees with the usual expression \cite{Yao2006Review-of-Parti} for the vacuum flavor oscillation probability as a function of spatial distance $L$ in a frame at rest with respect to the source and detector, as can be seen by noting that $p_\mathscr{W}^i = dx^i / d\lambda$ implies (in flat spacetime) that the worldline's affine parameter is equal to $\lambda_L = L / |{\bf p}_\mathscr{W} |$ when the neutrino has traveled a spatial distance $L$.

The operators giving rise to neutrino and antineutrino distribution matrices can be constructed from states $\neutrinoKet{(\lambda)}{\alpha}$ and $\antineutrinoKet{(\lambda)}{\alpha}$ respectively. The density operator corresponding to the pure state $\neutrinoKet{(\lambda)}{\alpha}$---that is, the operator whose matrix elements comprise the density matrix describing a single neutrino with worldline $\mathscr{W}$ that began life with definite flavor $\alpha$---is
\begin{equation}
\hat\rho_{\mathscr{W},\alpha}(\lambda) = \neutrinoKet{(\lambda)}{\alpha} \neutrinoBra{(\lambda)}{\alpha}. \label{singleParticleDensity}
\end{equation}
Suppose we have an ensemble of systems of noninteracting neutrinos with definite flavors $\alpha$ at $\lambda = 0$ on worldline $\mathscr{W}$. 
Let $f_\mathscr{W,\alpha}(0)$ be the ensemble-averaged number of $\alpha$ neutrinos at $\lambda = 0$ on $\mathscr{W}$; then the single-particle neutrino distribution operator describing the ensemble is
\begin{equation}
\hat\rho_\mathscr{W}(\lambda) = \sum_\alpha f_\mathscr{W,\alpha}(0) \, \hat\rho_{\mathscr{W},\alpha}(\lambda). \label{distributionOperator}
\end{equation}
Its equation of motion is
\begin{equation}
{\rm i}\frac{d}{d\lambda} \hat\rho_\mathscr{W}(\lambda) = \frac{1}{2} \left[ \hat M^2, \hat\rho_\mathscr{W}(\lambda)  \right], \label{operatorEquationOfMotion}
\end{equation}
which follows directly from Eqs. (\ref{schrodinger1}) and (\ref{singleParticleDensity}). With replacements $\neutrinoKet{(\lambda)}{\alpha} \rightarrow \antineutrinoKet{(\lambda)}{\alpha}$, $\hat\rho_{\mathscr{W},\alpha}(\lambda) \rightarrow \hat{\bar\rho}_{\mathscr{W},\alpha}(\lambda)$, $f_\mathscr{W,\alpha}(0) \rightarrow  \bar f_\mathscr{W,\alpha}(0)$, and $\hat\rho_{\mathscr{W}}(\lambda) \rightarrow \hat{\bar\rho}_{\mathscr{W}}(\lambda)$ the same construction holds, so that
\begin{equation}
{\rm i}\frac{d}{d\lambda} \hat{\bar\rho}_\mathscr{W}(\lambda) = \frac{1}{2} \left[ \hat M^2, \hat{\bar\rho}_\mathscr{W}(\lambda)  \right] \label{antineturinoOperatorEquationOfMotion}
\end{equation}
in the case of antineutrinos as well.

While the neutrino and antineutrino distribution operators obey the same equation of motion,
it is convenient to define the matrix representations of these equations in such a way that that they acquire a relative sign difference. The reason is that it is desirable for the matrix representations $M^2_{\alpha\beta}$ of the flavor-basis squared mass operator $\hat M^2$ to be the same in the neutrino and antineutrino cases (and equal to the square of the mass matrix in the Lagrangian for free neutrino flavor fields). In the neutrino case this requirement is consistent with the standard construction of a matrix representation:
\begin{equation}
\hat M^2 = \sum_{\alpha,\beta} M_{\alpha\beta}^2 \, \neutrinoKet{}{\alpha}\neutrinoBra{}{\beta},
\end{equation}
so that 
\begin{equation}
M^2_{\alpha\beta} = \neutrinoBra{}{\alpha} \hat M^2 \neutrinoKet{}{\beta}. \label{massElements1}
\end{equation}
However, to obtain the same representation $M^2_{\alpha\beta}$ in the antineutrino case the `backwards' definition
\begin{equation}
\hat M^2 = \sum_{\alpha,\beta} M_{\alpha\beta}^2 \, \antineutrinoKet{}{\beta}\antineutrinoBra{}{\alpha},
\end{equation}
so that
\begin{equation}
M^2_{\alpha\beta} = \antineutrinoBra{}{\beta} \hat M^2 \antineutrinoKet{}{\alpha}, \label{massElements2}
\end{equation}
is required to compensate for the opposite transformations of neutrino and antineutrino states in Eqs. (\ref{neutrinoBases}) and (\ref{antineutrinoBases}). If the neutrino and antineutrino distribution operators are expanded analogously,
\begin{eqnarray}
\hat\rho_\mathscr{W}(\lambda) &=& \sum_{\alpha,\beta} \rho_{\mathscr{W},\alpha\beta}(\lambda)\, \neutrinoKet{}{\alpha}\neutrinoBra{}{\beta} \label{neutrinoMatrixExpansion}, \\
\hat{\bar\rho}_\mathscr{W}(\lambda) &=& \sum_{\alpha,\beta} \bar\rho_{\mathscr{W},\alpha\beta}(\lambda)\, \antineutrinoKet{}{\beta}\antineutrinoBra{}{\alpha},
\end{eqnarray}
then the matrix representations $M^2$, $\rho_\mathscr{W}(\lambda)$, and $\bar\rho_\mathscr{W}(\lambda)$ all transform the same way in species (flavor/mass) space: if $A$ represents any of these matrices, then the flavor representations are related to the mass representations by $A_{\rm flavor} = U\, A_{\rm mass}\, U^\dagger$, where $U$ has elements $U_{\alpha i}$. 
However, the matrix representations of Eqs. (\ref{operatorEquationOfMotion}) and (\ref{antineturinoOperatorEquationOfMotion}) now acquire a sign difference:
\begin{eqnarray}
{\rm i}\frac{d}{d\lambda} \rho_\mathscr{W}(\lambda) &=& \frac{1}{2} \left[  M^2, \rho_\mathscr{W}(\lambda)  \right], \label{neutrinoEvolution}\\
{\rm i}\frac{d}{d\lambda} \bar\rho_\mathscr{W}(\lambda) &=& -\frac{1}{2} \left[  M^2, \bar\rho_\mathscr{W}(\lambda)  \right] \label{antineutrinoEvolution}
\end{eqnarray} 
for neutrinos and antineutrinos respectively. Note that the absence of hats indicates that these are matrix equations rather than operator equations.

The elements of  $\rho_\mathscr{W}(\lambda)$ and $\bar\rho_\mathscr{W}(\lambda)$ deserve further inspection. They are
\begin{eqnarray}
\rho_{\mathscr{W},\alpha\beta}(\lambda) &=& \sum_\gamma f_\mathscr{W,\gamma}(0) \neutrinoHalfBra{}{\alpha}\neutrinoKet{(\lambda)}{\gamma}  \nonumber \\
& &\times \neutrinoHalfBra{(\lambda)}{\gamma}\neutrinoKet{}{\beta}, \label{neutrinoElements}\\
\bar\rho_{\mathscr{W},\alpha\beta}(\lambda) &=& \sum_\gamma  f_\mathscr{W,\gamma}(0) \antineutrinoHalfBra{}{\beta}\antineutrinoKet{(\lambda)}{\gamma} \nonumber \\
& &\times \antineutrinoHalfBra{(\lambda)}{\gamma}\antineutrinoKet{}{\alpha}. \label{antineutrinoElements}
\end{eqnarray}
As the sum of the initial numbers of neutrinos in flavors $\gamma$ on worldline $\mathscr{W}$,  weighted by the probabilities of flavor transitions $\gamma \rightarrow \alpha$, the diagonal elements
\begin{equation}
\rho_{\mathscr{W},\alpha\alpha}(\lambda) = \sum_\gamma \left|\neutrinoHalfBra{}{\alpha}\neutrinoKet{(\lambda)}{\gamma}\right|^2\, f_\mathscr{W,\gamma}(0) 
\end{equation}
are equal to $f_\mathscr{W,\alpha}(\lambda)$, the number of neutrinos of flavor $\alpha$ at $\lambda$ (and similarly for antineutrinos).
Manifestly, the traces of $\rho_\mathscr{W}(\lambda)$ and $\bar\rho_\mathscr{W}(\lambda)$ are respectively equal to the numbers of neutrinos and antineutrinos of all species on worldline $\mathscr{W}$. The off-diagonal elements quantify the overlap in flavors $\alpha$ and $\beta$ generated from the initial numbers of neutrinos in flavors $\gamma$.

The desired Liouville equations are close at hand. A particular value of $\lambda$ on the worldline $\mathscr{W}$ specifies a point in spacetime, and the on-shell tangent vector to $\mathscr{W}$ coincides with the neutrino momentum. Therefore, if attention is broadened from a single worldline to a collection of them forming a congruence of curves in phase space, then the specification of $\mathscr{W}$ and dependence on $\lambda$ employed thus far are equivalent to dependence on $t, {\bf x}, {\bf p}$ in some coordinate system. Synchronize parametrizations in the congruence of curves such that for each worldline $\lambda = 0$ corresponds to $t=0$ in a chosen coordinate system; then the average particle numbers per worldline $f_\mathscr{W,\alpha}(0)$ and $f_\mathscr{W,\alpha}(\lambda)$ encountered above correspond to $f_\alpha(0,{\bf x},{\bf p})$ and $f_\alpha(t,{\bf x},{\bf p})$, where these latter quantities are classical distribution functions as in Eq. (\ref{classicalDistribution}). Therefore, with a choice of coordinate system, the $\rho_\mathscr{W}(\lambda)$ pertaining to a set of neighboring worldlines crossing an infinitesimal spacelike hypersurface in phase space may be denoted $\rho(t,{\bf x},{\bf p})\,d^3{\bf x}\, d^3{\bf p} / (2\pi)^3$, and similarly for antineutrinos. (Relativistic neutrinos and antineutrinos produced by $V-A$ interactions have spin degeneracy $g=1$).

Hence we have neutrino and antineutrino distribution matrices $\rho(t,{\bf x},{\bf p})$ and $\bar\rho(t,{\bf x},{\bf p})$; taking into account Eqs. (\ref{ddlambda}), (\ref{neutrinoEvolution}), and (\ref{antineutrinoEvolution}), together with the Liouville theorem (invariance of phase space volume elements \cite{Lindquist1966Relativistic-Tr,Ehlers1971General-Relativ,Israel1972The-Relativisti}), these distribution matrices satisfy the Liouville equations 
\begin{eqnarray}
p^\mu \frac{\partial}{\partial x^\mu} \rho(t,{\bf x},{\bf p}) + \frac{{\rm i}}{2} \left[  M^2, \rho(t,{\bf x},{\bf p}) \right] &=& 0, \label{neutrinoLiouville}\\
p^\mu \frac{\partial}{\partial x^\mu} \bar\rho(t,{\bf x},{\bf p}) -  \frac{{\rm i}}{2} \left[  M^2, \bar\rho(t,{\bf x},{\bf p}) \right] &=& 0. \label{antineutrinoLiouville}
\end{eqnarray}
The flavor/mass structure of the Hermitian matrices $M^2$, $\rho(t,{\bf x},{\bf p})$, and $\bar\rho(t,{\bf x},{\bf p})$ is given in Eqs. (\ref{massElements1}), (\ref{massElements2}) and (\ref{neutrinoElements}), (\ref{antineutrinoElements}), and their transformation properties are described in the text between these. The diagonal elements of the distribution matrices are real and are classical distribution functions, as in Eq. (\ref{classicalDistribution}), for the particle types of the chosen representation (flavor or mass). The off-diagonal elements quantify the extent of mixing (species superpositions) present in the neutrino gas.

Observable neutrino flavor mixing phenomena depend on the differences of squared mass eigenvalues $\delta_{ji} \equiv m_j^2 - m_i^2$, and (aside from upper limits) such differences are the only data on neutrino mass that have been experimentally determined  \cite{Yao2006Review-of-Parti}. That flavor mixing probabilities do not depend on absolute masses (other than satisfaction of the relativistic limit) is apparent when the squared mass matrix is decomposed as $M^2 = \Sigma + \Delta$, where $\Sigma$ is proportional to the identity matrix and $\Delta$ is the traceless part. For instance, in the standard case of three neutrino species,
\begin{eqnarray}
(\Sigma) &=& \frac{1}{3}\,{\rm Tr}(M^2)  \\
&=& \frac{1}{3}\left(m_1^2+m_2^2+m_3^2\right)\pmatrix{1 & 0 &0 \cr 0 & 1 & 0 \cr 0 & 0 & 1}
\end{eqnarray}
in any basis, and
\begin{eqnarray}
(\Delta)_{\rm mass}\!\!\! &=&\!\!\! (M^2)_{\rm mass} - (\Sigma) \\
\!\!\! &=&\!\!\! \frac{1}{3}\pmatrix{-\delta_{21} -\delta_{31}& 0 &0 \cr 0 & \delta_{21} -\delta_{32} & 0 \cr 0 & 0 & \delta_{32} +\delta_{31}} \label{massDifferenceMatrix}
\end{eqnarray}
in the mass basis. Because $\Sigma$ cancels out of the commutator, Eqs. (\ref{neutrinoLiouville})-(\ref{antineutrinoLiouville}) become \footnote{In curved spacetime, or with use of curvilinear coordinates and/or an accelerated reference frame for reckoning momenta, momentum derivative terms would arise from the relation $d/d\lambda = p^\mu\, \partial/\partial x^\mu - {\Gamma^\mu}_{\nu\rho} p^\nu p^\rho \, \partial/\partial p^\mu$ (see for instance Refs. \cite{Cardall2003Conservative-fo,Cardall2005Conservative-sp}). In addition, if there are interactions that flip neutrino spin (such as the action of a magnetic field upon a neutrino magnetic moment), then the effects of a `spin connection' resulting from curved spacetime must be taken into account \cite{Cardall1997Neutrino-oscill}.}
\begin{eqnarray}
p^\mu \frac{\partial}{\partial x^\mu} \rho(t,{\bf x},{\bf p}) + \frac{{\rm i}}{2} \left[  \Delta, \rho(t,{\bf x},{\bf p}) \right] &=& 0, \label{neutrinoLiouville2}\\
p^\mu \frac{\partial}{\partial x^\mu} \bar\rho(t,{\bf x},{\bf p}) -  \frac{{\rm i}}{2} \left[ \Delta, \bar\rho(t,{\bf x},{\bf p}) \right] &=& 0. \label{antineutrinoLiouville2}
\end{eqnarray}
And going back to Eq. (\ref{schrodinger1}), $\Sigma$ merely gives rise to an overall phase that cancels in flavor transition probabilities.

\section{Starting from a quantum `density function'}
\label{sec:correlation}

A more fundamental approach than that presented in the previous section begins with a quantum `density function,' the mean value of a pair of normal-ordered neutrino quantum field operators. (Unlike the previous section, the convention of denoting operators by hats is abandoned here because distinctions between operators and matrix representations thereof using the same symbol will not be required.) The focus here is on free neutrino fields. Some details regarding these---including conventions, and reminders about behavior in the relativistic limit---are given in the Appendix. 

\subsection{Quantum density function}

Define the free-field quantum `density function' $\Gamma(y,z)$---a function of spacetime positions $y$ and $z$---in terms of the neutrino quantum field operators $\nu(y)$ and $\bar\nu(z)$:
\begin{equation}
{\rm i}\,\Gamma^{\ell m}_{ij}(y,z) = \left\langle N \nu_i^\ell(y)\, \bar \nu_j^m(z) \right\rangle. \label{densityFunction}
\end{equation}
The subscripts $i,j$ index fields of definite mass, which are the `physical fields' for which the usual quantization in terms of Fock states makes sense \cite{Giunti1992Remarks-on-the-}. The superscripts $\ell, m$ are spinor indices. 
In this case the bar on $\bar \nu(z)$ denotes the Pauli conjugate $\bar \nu(z) = \nu^\dagger(z)\, \gamma^0$; note that in most other instances later in this section a bar simply labels a quantity related to antineutrinos rather than a Pauli conjugate. The angle brackets signify both the taking of an expectation value with respect to a many-particle quantum state and an average over a statistical ensemble of such quantum states. The $N$ on the right-hand side denotes `normal ordering,' which specifies that creation operators are to be placed to the left of annihilation operators, with the introduction of minus signs appropriate to the interchange of fermionic operators as needed. In particular, separating the neutrino field operator 
\begin{equation}
\nu(y) = A(y) + B(y)
\end{equation} 
into its positive- and negative-frequency parts $A(y)$ and $B(y)$, the density operator becomes
\begin{equation}
{\rm i}\,\Gamma^{\ell m}_{ij}(y,z) = -\left\langle \bar A_j^m(z) A_i^\ell(y) \right\rangle + \left\langle B_i^\ell(y) \bar B_j^m(z) \right\rangle. \label{densityFunction2}
\end{equation}
As will become clear below, the first and second terms are associated with the densities of neutrinos and antineutrinos respectively.
Rapidly oscillating cross terms between positive- and negative-frequency parts are not relevant to the macroscopic limit, and have been dropped \cite{Sigl1993General-kinetic,Groot1980Relativistic-Ki}.
 
In most cases of practical interest the complications of spin can be eliminated. This is because the combination of $V-A$ neutrino interactions with a relativistic limit to ${\cal O}(m_\nu^2/ E_\nu)$,   where $m_\nu$ and $E_\nu$ are characteristic neutrino mass and energy scales, ensures that only negative-helicity neutrinos and positive-helicity antineutrinos need be considered. The simplifications that result from taking spin transitions off the table are twofold. 

The first simplification resulting from the irrelevance of spin pertains to the equations of motion employed. While $\Gamma^{\ell m}_{ij}(y,z)$ obeys the Dirac equation by virtue of its construction from Dirac fields, each spinor-space component of this density function also obeys the Klein-Gordon equation. Hence, when considerations of spin are irrelevant, the Klein-Gordon equation can be used from the outset. At first glance this may seem counterproductive, for the same reason the Dirac equation was invented in the first place: like Dirac---and with a somewhat related motivation, namely the maintenance of positive probability distributions---we are ultimately after equations that are first order rather than second order in time. However, we shall see that in the present context the desired first-order equations emerge very naturally from a combination of Klein-Gordon equations, by virtue of a Wigner transformation.
 
The second simplification resulting from the neglect of spin is that the $4\times 4$ spinor structure of  $\Gamma^{\ell m}_{ij}(y,z)$ can be eliminated, so that focus shifts to entities without spinor indices.

\subsection{Obtaining a first-order equation}
\label{subsec:firstOrder}

An explicit account of these simplifications begins with two Klein-Gordon equations satisfied by the density function before the relativistic limit is taken---one with respect to $y$, and one with respect to $z$. Written in matrix form with species indices suppressed, these two equations are
\begin{eqnarray}
\boxempty_y \, \Gamma^{\ell m}(y,z) + M^2 \,  \Gamma^{\ell m}(y,z)&=& 0, \label{kleingordon1}\\
\boxempty_z \, \Gamma^{\ell m}(y,z) +  \Gamma^{\ell m}(y,z) \, M^2 &=& 0, \label{kleingordon2}
\end{eqnarray}
where (for instance) 
\begin{equation}
\boxempty_y \equiv \frac{\partial}{ \partial y_\mu} \frac{\partial}{\partial y^\mu} = \frac{\partial^2}{(\partial y^0)^2} - \nabla_{\bf y}^2
\end{equation}
is the d'Alembertian with respect to spacetime position $y$.
The difference of Eqs. (\ref{kleingordon1}) and (\ref{kleingordon2}) is
\begin{equation}
\left(\boxempty_y - \boxempty_z\right)  \Gamma^{\ell m}(y,z) + \left[ \Delta,  \Gamma^{\ell m}(y,z) \right] = 0, \label{kleinGordonDifference}
\end{equation}
in which $M^2$ has been replaced by its traceless part $\Delta$ containing only squared mass differences as described in the last paragraph of Sec. \ref{sec:simple}.

Turn now to the change of variables associated with a Wigner transformation. Rewrite $y$ and $z$ in terms of new spacetime variables $x$ and $\Xi$:
\begin{eqnarray}
y &=& x + \frac{\Xi}{2}, \label{forwardChange1} \\
z &=& x - \frac{\Xi}{2}. \label{forwardChange2}
\end{eqnarray}
The meanings of $x$ and $\Xi$ begin to become more apparent from the inverse transformations
\begin{eqnarray}
x &=& \frac{1}{2}(y + z), \label{backwardChange1} \\
\Xi &=& y-z.
\end{eqnarray}
The average spacetime position $x$ will become the `macroscopic' position variable in classical expressions obtained from the present quantum formalism. The difference coordinate $\Xi$ is indirectly related to the `macroscopic' momentum variable in such classical expressions. In particular, the Wigner transformation is a Fourier transformation with respect to $\Xi$. The Wigner transformation of $\Gamma^{\ell m}_{ij}(y,z)$ yields the `mixed representation' $\mathcal{G}^{\ell m}_{ij}(x, P)$ of the density function:
\begin{equation}
\mathcal{G}^{\ell m}_{ij}(x, P) = \int d^4 \Xi \; \mathrm{e}^{{\rm i} P\cdot \Xi} \;  \Gamma^{\ell m}_{ij}\left(x+\frac{\Xi}{2}, x - \frac{\Xi}{2}\right). \label{neutrinoWigner}
\end{equation}
At this point $P$ is not an on-shell four-momentum. However we shall see below that it does basically become the `macroscopic' momentum variable in derived classical expressions (up to a sign difference between the neutrino and antineutrino parts).

The commutator in Eq. (\ref{kleinGordonDifference}) already looks familiar from the Liouville equations obtained at the end of Sec. \ref{sec:simple}; it turns out that the change of variables of Eqs. (\ref{forwardChange1}) and (\ref{forwardChange2}) associated with the Wigner transformation starts to bring the differential operator term into more familiar form as well. Under this change of variables the second-order operator becomes
\begin{equation}
\boxempty_y - \boxempty_z = 2\, \frac{\partial}{\partial \Xi} \cdot \frac{\partial}{\partial x}, 
\label{differentialOperator}
\end{equation}
which is first order with respect to $x$. Note also that
\begin{eqnarray}
 \Gamma^{\ell m}_{ij}(y,z) &=&  \Gamma^{\ell m}_{ij}\left(x+\frac{\Xi}{2}, x - \frac{\Xi}{2}\right) \\
&=& \int \frac{d^4 P}{(2\pi)^4} \; \mathrm{e}^{-{\rm i} P\cdot \Xi} \; \mathcal{G}^{\ell m}_{ij}(x, P), \label{inverseWigner} 
\end{eqnarray}
where the transformation in the second line is the inverse of that in Eq. (\ref{neutrinoWigner}). 
The mixed representation of the density function satisfies
\begin{equation}
-2\,{\rm i} \, P^\mu \frac{\partial}{\partial x^\mu} \, \mathcal{G}^{\ell m}(x,P) + \left[ \Delta,   \mathcal{G}^{\ell m}(x,P) \right] = 0, \label{neutrinoTransformedKleinGordon}
\end{equation}
which follows from substitution of Eqs. (\ref{differentialOperator}) and (\ref{inverseWigner}) into Eq. (\ref{kleinGordonDifference}).

\subsection{Interpretation of the mixed representation}
\label{subsec:interpretation}

Consider next the spinor-space structure of the density function in the case of practical interest described above, in which the neutrino and antineutrino populations are overwhelmingly relativistic. In this case it follows from the explicit expressions in the Appendix that the spinor space structure of the density function reduces to
\begin{eqnarray}
\left(  \Gamma^{\ell m}(y,z) \right) &\rightarrow& \pmatrix{ 0 &   \Gamma^{LR}(y,z) \cr 0 & 0 }.
\end{eqnarray}
Note that in the relativistic limit only one $2\times 2$ block is nonzero. The notation $ \Gamma^{LR}(y,z)$ for this block denotes the fact that it would be the block projected out if $ \Gamma(y,z)$ were sandwiched between the left- and right-projection matrices $P_L$ and $P_R$ of Eqs. (\ref{projectionLeft}) and (\ref{projectionRight}). Explicitly, 
\begin{eqnarray}
{\rm i}\, \Gamma^{LR}_{ij}(y,z) &=& - \int  \frac{d^3{\bf q}}{(2\pi)^3} \frac{d^3{\bf u}}{(2\pi)^3}  \left[\mathrm{e}^{{\rm i} (u_j\cdot z - q_i\cdot y)} \mathrm{N}_{ij}^{LR}({\bf q},{\bf u}) \right.\nonumber \\
& & \left. \ \ \ \ \ \  - \  \mathrm{e}^{{\rm i} (q_i\cdot y - u_j\cdot z)} \bar \mathrm{N}_{ij}^{LR}({\bf q},{\bf u})\right].\label{neutrinoLRcoordinate} 
\end{eqnarray}
Here $(q_i^\mu)=(E_{{\bf q},i},{\bf q})$ with $E_{{\bf q},i} \equiv \sqrt{|{\bf q}|^2 + m_i^2}\approx |{\bf q}| + m_i^2/2|{\bf q}|$ (and similarly for the components of $u_j$), and
to ${\cal O}(m_\nu^2/ E_\nu)$ 
\begin{eqnarray}
\mathrm{N}_{ij}^{LR}({\bf q},{\bf u}) &=& \xi_{\bf q}^\downarrow \, {\xi_{\bf u}^\downarrow}^\dagger \, \langle a_{{\bf u}, \downarrow,j}^\dagger a_{{\bf q}, \downarrow,i}\rangle, \label{relativisticNeutrinoExpectationValue0}\\
\bar \mathrm{N}_{ij}^{LR}({\bf q},{\bf u}) &=& \eta_{\bf q}^\uparrow \, {\eta_{\bf u}^\uparrow}^\dagger \, \langle b_{{\bf q}, \uparrow,i}^\dagger b_{{\bf u}, \uparrow,j}\rangle  \label{relativisticAntineutrinoExpectationValue0},
\end{eqnarray}
as discussed in the Appendix. (Note that $\xi_{\bf q}^\downarrow$ and $\eta_{\bf q}^\uparrow$ are two-component spinors.)

Moving to the mixed representation provides a convenient means of separating the neutrino and antineutrino parts of the density function, by making manifest its positive- and negative-frequency parts.
Apply the Wigner transformation of Eq. (\ref{neutrinoWigner}) to Eq. (\ref{neutrinoLRcoordinate}) and find
\begin{eqnarray}
{\rm i}\, \mathcal{G}^{LR}_{ij}(x,P) &=& - \int  \frac{d^3{\bf q}}{(2\pi)^3} \frac{d^3{\bf u}}{(2\pi)^3}  \left[\mathrm{e}^{-{\rm i}(q_i-u_j)\cdot x} \mathrm{N}_{ij}^{LR}({\bf q},{\bf u}) \right.\nonumber \\
& & \ \ \ \ \ \ \ \ \ \ \times\, (2\pi)^4\,\delta^4\left(P - \frac{q_i+u_j}{2}\right) \nonumber \\
& & \left. \ \ \ \ \ \  - \  \mathrm{e}^{{\rm i}(q_i-u_j)\cdot x} \bar \mathrm{N}_{ij}^{LR}({\bf q},{\bf u})\right. \nonumber \\
& & \left. \ \ \ \ \ \ \ \ \ \ \times\, (2\pi)^4\,\delta^4\left(P + \frac{q_i+u_j}{2}\right)\right].\label{neutrinoLRmixed} 
\end{eqnarray}
Because $q_i$ and $u_j$ are on-shell momenta, it is evident that $P^0 > 0$ in the first term and $P^0 < 0$ in the second term. Separate neutrino and antineutrino density functions are obtained by projecting out these positive- and negative-frequency parts with step functions $\theta\left(P^0\right)$ and $\theta\left(-P^0\right)$:
\begin{eqnarray}
{\rm i}\, G^{LR}_{ij}(x,p) &=& {\rm i}\int d^4 P \; \delta^4\left(p- P\right) \, \theta\left(P^0\right) \, \mathcal{G}^{LR}_{ij}(x,P) \nonumber \\
&=& - \int  \frac{d^3{\bf q}}{(2\pi)^3} \frac{d^3{\bf u}}{(2\pi)^3}  \mathrm{e}^{-{\rm i}(q_i-u_j)\cdot x} \mathrm{N}_{ij}^{LR}({\bf q},{\bf u}) \nonumber \\
& & \ \ \ \ \ \ \ \ \ \ \times\, (2\pi)^4\,\delta^4\left(p - \frac{q_i+u_j}{2}\right) \label{neutrinoDensityFunction}
\end{eqnarray}
and
\begin{eqnarray}
{\rm i}\, \bar G^{LR}_{ij}(x,p) &=& {\rm i}\int d^4 P \; \delta^4\left(p+ P\right) \, \theta\left(-P^0\right) \, \mathcal{G}^{LR}_{ij}(x,P) \nonumber \\
&=& \int  \frac{d^3{\bf q}}{(2\pi)^3} \frac{d^3{\bf u}}{(2\pi)^3}  \mathrm{e}^{{\rm i}(q_i-u_j)\cdot x} \bar \mathrm{N}_{ij}^{LR}({\bf q},{\bf u}) \nonumber \\
& & \ \ \ \ \ \ \ \ \ \ \times\, (2\pi)^4\,\delta^4\left(p - \frac{q_i+u_j}{2}\right). \label{antineutrinoDensityFunction}
\end{eqnarray}
Note that delta functions included in the projection operations yield the momentum label changes $P\rightarrow p$ in the case of the neutrino density function $G^{LR}_{ij}(x,p)$ and $P\rightarrow -p$ in the case of the antineutrino density function $\bar G^{LR}_{ij}(x,p)$. 
These density functions satisfy
\begin{eqnarray}
-2\,{\rm i} \, p^\mu \frac{\partial}{\partial x^\mu} \, G^{LR}(x,p) + \left[ \Delta, G^{LR}(x,p) \right] &=& 0, \label{neutrinoDensityEquation} \\
2\,{\rm i} \, p^\mu \frac{\partial}{\partial x^\mu} \, \bar G^{LR}(x,p) + \left[ \Delta, \bar G^{LR}(x,p) \right] &=& 0, \label{antineutrinoDensityEquation} 
\end{eqnarray}
which follow from applying to Eq. (\ref{neutrinoTransformedKleinGordon}) the same projections appearing in Eqs. (\ref{neutrinoDensityFunction}) and (\ref{antineutrinoDensityFunction}). 

In general cases the mixed representation provides {\em complementary} position and momentum probability distributions, as required by the quantum mechanical incompatibility of position and momentum, rather than a {\em joint} position/momentum distribution.

\subsubsection{Position distribution}

A position distribution is obtained by integrating the mixed representation over all `momenta.' This can be seen by comparison of the diagonal elements of Eqs. (\ref{neutrinoDensityFunction}) and (\ref{antineutrinoDensityFunction}) in species space with the relevant component of the (normal-ordered) number current $\langle N \bar\nu_i(x) \gamma^\mu \nu_i(x)\rangle$ of species $i$. 

In particular, the net neutrino number density of species $i$---that is, the difference between the neutrino and antineutrino number densities or position distributions $n_i(x)$ and $\bar n_i(x)$---is the 0th spacetime component of the number current:
\begin{eqnarray}
n_i(x) - \bar n_i(x) &=& \left\langle N \bar\nu_i(x) \gamma^0 \nu_i(x)\right\rangle.
\end{eqnarray}
More explicitly,
\begin{eqnarray}
n_i(x) - \bar n_i(x) &=& \int  \frac{d^3{\bf q}}{(2\pi)^3} \frac{d^3{\bf u}}{(2\pi)^3}  \left[\mathrm{e}^{-{\rm i}(q_i-u_i)\cdot x} \mathcal{N}_i({\bf q},{\bf u}) \right.\nonumber \\
& & \left. \ \ \ \ \ \  - \  \mathrm{e}^{{\rm i}(q_i-u_i)\cdot x} \bar \mathcal{N}_i({\bf q},{\bf u})\right], \label{numberDensity} 
\end{eqnarray}
where
\begin{eqnarray}
\mathcal{N}_i({\bf q},{\bf u}) &=&  {\xi_{\bf u}^\downarrow}^\dagger \, \xi_{\bf q}^\downarrow \, \langle a_{{\bf u}, \downarrow,i}^\dagger a_{{\bf q}, \downarrow,i}\rangle, \label{relativisticNeutrinoExpectationValue1}\\
\bar \mathcal{N}_i({\bf q},{\bf u}) &=& {\eta_{\bf u}^\uparrow}^\dagger \, \eta_{\bf q}^\uparrow  \, \langle b_{{\bf q}, \uparrow,i}^\dagger b_{{\bf u}, \uparrow,i}\rangle  \label{relativisticAntineutrinoExpectationValue1}.
\end{eqnarray}
Aside from the single species index $i$ compared with the potentially different indices $i$ and $j$, the difference between Eqs. (\ref{relativisticNeutrinoExpectationValue1}), (\ref{relativisticAntineutrinoExpectationValue1}) and (\ref{relativisticNeutrinoExpectationValue0}), (\ref{relativisticAntineutrinoExpectationValue0}) is that the one pair has inner products of two-component spinors, while the other has an outer product giving rise to a $2\times 2$ matrix in spinor space. 

The neutrino and antineutrino contributions to Eq. (\ref{numberDensity}) are readily obtained from the species-space diagonal components of Eqs. (\ref{neutrinoDensityFunction}) and (\ref{antineutrinoDensityFunction}) respectively. Integrating over $p$, and using the fact that the inner product of any two spinors is equal to the trace of their outer product, one finds that 
\begin{equation}
n_i(x) - \bar n_i(x) = - \int \frac{d^4 p}{(2\pi)^4}\, \mathrm{Tr}\left[{\rm i}\, G^{LR}_{ii}(x,p) + {\rm i}\, \bar G^{LR}_{ii}(x,p) \right],
\end{equation}
where the trace is over the spinor indices of the $2\times 2$ blocks. This motivates the definition of spatial `number density matrices'
\begin{equation}
\rho_{ij}(t,{\bf x}) = \int \!\frac{d^4 p}{(2\pi)^4}\;{\rm Tr}\left[-\mathrm{i}  \, G^{LR}_{ij}(t,{\bf x},p)\right], \label{neutrinoSpatialDensity}
\end{equation}
and
\begin{equation}
\bar\rho_{ij}(t,{\bf x}) = \int \!\frac{d^4 p}{(2\pi)^4}\;{\rm Tr}\left[\mathrm{i}  \, \bar G^{LR}_{ij}(t,{\bf x},p)\right], \label{antineutrinoSpatialDensity}
\end{equation}
whose diagonal elements are spatial number densities of the various neutrino and antineutrino species respectively. (The dependence on spacetime components $(x^\mu)=(t,{\bf x})$ has been displayed here more explicitly.)

\subsubsection{Momentum distribution}

A momentum distribution is obtained by integrating the mixed representation over the volume of the system. Integrating Eq. (\ref{numberDensity}) over all space results in a factor $(2\pi)^3 \delta^3({\bf u} - {\bf q})$, so that the net total neutrino number of species $i$ is
\begin{eqnarray}
N_i - \bar N_i &=& \int d^3{\bf x} \left[ n_i(x) - \bar n_i(x) \right] \nonumber \\
&=& \int \frac{d^3{\bf q}}{(2\pi)^3} \left( \langle a_{{\bf q}, \downarrow,i}^\dagger a_{{\bf q}, \downarrow,i}\rangle - \langle b_{{\bf q}, \uparrow,i}^\dagger b_{{\bf q}, \uparrow,i}\rangle \right). \nonumber \\
& & \label{totalNumber}
\end{eqnarray}
Comparison with the definition of an occupation number in Eq. (\ref{occupationNumber})---whose integral over ${\bf q}$ also gives a total number of particles in the system---implies that the difference of neutrino and antineutrino occupation numbers or momentum distributions is
\begin{equation}
n_i(t,\mathbf{q}) - \bar n_i(t,\mathbf{q}) = \frac{1}{V}\left( \langle a_{{\bf q}, \downarrow,i}^\dagger a_{{\bf q}, \downarrow,i}\rangle - \langle b_{{\bf q}, \uparrow,i}^\dagger b_{{\bf q}, \uparrow,i}\rangle \right).
\end{equation}
Precisely the same expression, but with momentum label $\mathbf{q}$ replaced by $\mathbf{p}$, is obtained from the diagonal elements of Eqs. (\ref{neutrinoDensityFunction}) and (\ref{antineutrinoDensityFunction}) by integrating over ${\bf x}$, integrating over $p^0$ (which puts the momentum $p$ on shell), and making use of Eqs. (\ref{qNeutrinoExpectationValueTrace}) and (\ref{qAntineutrinoExpectationValueTrace}) in the Appendix:
\begin{eqnarray}
\lefteqn{n_i(t,\mathbf{p}) - \bar n_i(t,\mathbf{p})} \nonumber \\
 &=& - \frac{1}{V}\int d^3{\bf x} \int \frac{d p^0}{(2\pi)}\, \mathrm{Tr}\left[{\rm i}\, G^{LR}_{ii}(x,p) + {\rm i}\, \bar G^{LR}_{ii}(x,p) \right]. \nonumber \\
 & &
\end{eqnarray}
This motivates the definition of `occupation matrices'
\begin{equation}
\rho_{ij}(t,{\bf p}) = \frac{1}{V}\int d^3{\bf x}  \int \!\frac{dp^0}{2\pi}\;{\rm Tr}\left[-{\rm i} \, G^{LR}_{ij}(t,{\bf x},p)\right]
\end{equation}
and
\begin{equation}
\bar\rho_{ij}(t,{\bf p}) = \frac{1}{V}\int d^3{\bf x} \int \!\frac{dp^0}{2\pi} \;{\rm Tr}\left[{\rm i} \,\bar G^{LR}_{ij}(t,{\bf x},p)\right],
\end{equation}
whose diagonal elements are occupation numbers of the various neutrino and antineutrino species respectively. (Here the components of ${\bf p}={\bf q}={\bf u}$ are quantum numbers of momentum eigenstates, a role denoted by ${\bf q}$ in Sec. \ref{sec:introduction}.)

\subsubsection{Classical joint position/momentum distribution}

Beyond the complementary position and momentum probability distributions generally available from the mixed representation of a density function, a joint position/momentum probability distribution---akin to a classical one-particle distribution function or phase space density---is expected to emerge when the spacetime and momentum dependences of a neutrino ensemble satisfy appropriate conditions. In particular a classical system is characterized by
\begin{equation}
E T \gg 1, \ \ \ P L \gg 1, \label{classicalityConditions}
\end{equation}
where $E$ and $P$ are characteristic energy and momentum scales and $T$ and $L$ are characteristic time and length scales.  

In order to elucidate this classical limit it is necessary to examine $\mathrm{N}_{ij}^{LR}({\bf q},{\bf u})$ and $\bar \mathrm{N}_{ij}^{LR}({\bf q},{\bf u})$ in Eqs. (\ref{neutrinoDensityFunction}) and (\ref{antineutrinoDensityFunction}) more closely, beginning with a specification of the state with respect to which the expectation values in Eqs. (\ref{relativisticNeutrinoExpectationValue0}) and (\ref{relativisticAntineutrinoExpectationValue0}) are taken. For illustrative purposes a pure state $\left| \Phi_N\right\rangle$ of $N$ neutrinos will be discussed here. The extension to systems represented by pure or mixed states which also include antineutrinos or other particle types might provoke complications (or at least changes) in notation but would involve no significant additional conceptual difficulties.

A pure state $\left| \Phi_N\right\rangle$ of $N$ neutrinos is built up out of multiparticle momentum eigenstates $\left| {\bf k}_1 i_1 \dots {\bf k}_N i_N\right\rangle$, where the ${\bf k}$ are momenta and the $i$ label mass eigenvalues.  These momentum eigenstates result from the action of antisymmetrized products of neutrino creation operators $a^\dagger_{{\bf k},\downarrow,i}$ upon the vacuum, together with energy factors $\sqrt{2 E_{{\bf k}}}$, such that these antisymmetric states are normalized according to
\begin{eqnarray}
\lefteqn{\left\langle  {\bf k}'_1 i'_1 \dots {\bf k}'_{N'} i'_{N'} | {\bf k}_1 i_1 \dots {\bf k}_N i_N\right\rangle } \nonumber \\
&=&\!\! \delta_{N'N} \!\!\sum_{\cal P}\! \delta_{\cal P}\! \prod_{a=1}^N \! (2 E_{{\bf k}_a})(2\pi)^3 \delta^3({\bf k}'_{{\cal P}a} - {\bf k}_a) \delta_{i'_{{\cal P}a} i_a}. \label{eigenstateNormalization}
\end{eqnarray}
The sum is over all permutations ${\cal P}$ of the list of particle labels $1\dots N$ indexed by $a$, with $\delta_{\cal P}$ equal to $1$ for even permutations and $-1$ for odd permutations, while ${\cal P}a$ is the particle label moved to the $a$th position under a particular permutation ${\cal P}$. This sum reflects the complete antisymmetry of these multi-neutrino states in that it vanishes if any two momentum/mass label pairs are equal.

In accordance with the assumption that the neutrino gas can be represented by a density function---that is, that correlations can be ignored---take $\left| \Phi_N\right\rangle$ to be a superposition of momentum eigenstates constructed with $N$ independent single-particle wave packets: 
\begin{eqnarray}
\left| \Phi_N\right\rangle &=& \left(\prod_{a=1}^N \int \frac{d^3{\bf k}_a}{(2\pi)^3} \frac{\phi_{{\bf p}_a}({\bf k}_a)}{\sqrt{2 E_{{\bf k}_a}}} \mathrm{e}^{-\mathrm{i}{\bf k}_a \cdot {\bf x}_{a,0}} \sum_{i_a} U_{\alpha_a i_a}^* \right)\nonumber \\
& & \times \left| {\bf k}_1 i_1 \dots {\bf k}_N i_N\right\rangle. \label{multineutrinoState}
\end{eqnarray}
The $\phi_{\bf p}({\bf k})$ are real-valued wave packet `envelopes' centered on ${\bf p}$, each normalized such that
\begin{equation}
\int \frac{d^3{\bf k}}{(2\pi)^3}\, [ \phi_{{\bf p}}({\bf k}) ]^2 = 1. \label{momentumNormalization}
\end{equation}
According to the usual wave packet technology these momentum-space envelopes are related to position-space wave packet envelopes $\psi_{{\bf x}_0,{\bf p}}({\bf x})$, peaked about ${\bf x}_0$, by
\begin{equation}
\psi_{{\bf x}_0,{\bf p}}({\bf x}) = \int \frac{d^3{\bf k}}{(2\pi)^3} \, \phi_{{\bf p}}({\bf k}) \, \mathrm{e}^{\mathrm{i}{\bf k} \cdot ({\bf x}-{\bf x}_{0})},
\end{equation}
whose normalization
\begin{equation}
\int d^3{\bf x}\, | \psi_{{\bf x}_0,{\bf p}}({\bf x}) |^2 = 1 \label{positionNormalization}
\end{equation}
follows from Eq. (\ref{momentumNormalization}). (For instance, if the $\phi_{\bf p}({\bf k})$ are taken to be Gaussian, then
\begin{equation}
\psi_{{\bf x}_0,{\bf p}}({\bf x}) =  \psi_{{\bf x}_0}({\bf x}) \, \mathrm{e}^{\mathrm{i}\mathbf{p} \cdot (\mathbf{x}-\mathbf{x}_0)}, 
\end{equation}
where $\psi_{{\bf x}_0}({\bf x})$ is a (real-valued) Gaussian centered on ${\bf x}_0$.)

Before explaining the particular choice of superposition of mass eigenstates represented by the sum over $i_a$ in Eq. (\ref{multineutrinoState}) it is helpful to clarify that $\left| \Phi_N\right\rangle$ is a Heisenberg-picture state which for definiteness is taken to correspond to a collection of $N$ relativistic neutrinos at $t=0$. These neutrinos are somewhat localized in both momentum space and position space, the $a$th neutrino being localized around ${\bf p}_a$ and ${\bf x}_{0,a}$ respectively (where the subscript $0$ is a reminder that this position localization is that which applies at $t = 0$). Hence the particular superposition $\sum_{i_a} U_{\alpha_a i_a}^*$ applies to a system in which the $a$th neutrino is in `definite flavor' $\alpha_a$ at $t=0$. This corresponds to the assumption used for the sake of illustration in Sec. \ref{sec:simple} that all the neutrinos were in definite flavors $\alpha$ at $t=0$. More generally the various neutrinos could be taken to be have been created in definite flavors (that is, in association with different charged leptons) at various different times $t_a < 0$; in this case a more general superposition $\sum_{i_a} c_{\alpha_a i_a}$ with $\sum_{i_a} |c_{\alpha_a i_a}|^2=1$ would apply, where $c_{\alpha_a i_a}$ encodes the amplitude for the $a$th neutrino to be in mass eigenstate $i_a$ after evolution from its creation in flavor $\alpha_a$ at $t_a<0$ until $t=0$.  

In order to obtain expectation values with respect to $\left| \Phi_N\right\rangle$ it is necessary to know its norm, and this in turn requires an understanding of how the Pauli exclusion principle---manifest in the complete antisymmetry of the multiparticle momentum eigenstates $\left| {\bf k}_1 i_1 \dots {\bf k}_N i_N\right\rangle$---impacts our localized particles. Consider a situation in which (at least) two neutrinos have the same flavor content, and also wave packet envelopes $\phi_{\bf p}({\bf k})$ in Eq. (\ref{multineutrinoState}) with identical shapes, centroids ${\bf p}$, and spatial offsets ${\bf x}_0$. In this case $\left| \Phi_N\right\rangle$ vanishes, because a product of wave packet envelopes/mass amplitudes that is symmetric with respect to (at least) two sets of quantum numbers ${\bf k}, i$ is `contracted' with the completely antisymmetric state $\left| {\bf k}_1 i_1 \dots {\bf k}_N i_N\right\rangle$. More generally, the norm of $\left| \Phi_N\right\rangle$ is
\begin{eqnarray}
\lefteqn{\left\langle \Phi_N | \Phi_N\right\rangle = 1 }\nonumber \\
&+&  \sum_{{\cal P}\ne 1} \delta_{\cal P} \prod_{a=1}^N  \delta_{\alpha_{{\cal P}a} \alpha_a} \int \frac{d^3{\bf k}_a}{(2\pi)^3} \, \phi_{{\bf p}_{{\cal P}a}}({\bf k}_a) \, \phi_{{\bf p}_a}({\bf k}_a) \nonumber \\
& &\times \, \mathrm{e}^{\mathrm{i}{\bf k}_a \cdot ({\bf x}_{{\cal P}a,0} - {\bf x}_{a,0})}, \label{fullNorm}
\end{eqnarray}
which follows from Eqs. (\ref{eigenstateNormalization})-(\ref{momentumNormalization}) and the unitarity of the mixing matrix. The sum is now over all permutations except the identity. In the aforementioned case of perfect wave packet overlap this sum becomes $-1$, and $\left\langle \Phi_N | \Phi_N\right\rangle$ vanishes. At the other extreme, if there is absolutely no wave packet overlap in either position or momentum space, then the sum in second and third lines of Eq. (\ref{fullNorm}) vanishes, so that $\left\langle \Phi_N | \Phi_N\right\rangle$ is unity. Between these extremes partial wave packet overlaps give this sum a value somewhere between $0$ and $-1$, and in turn the norm takes continuous values between $1$ and $0$.  Therefore  `wave packet smearing' has a graduated impact upon manifestations of the exclusion principle in `microscopic' representations of localized particles. 

However, with an end goal of a macroscopic treatment of a neutrino gas it is neither possible nor desirable to follow the details of all particles' wave packet shapes and overlaps; what is of interest instead is how the Pauli exclusion principle percolates down to the classical limit. To this end take a view that is sufficiently `coarse-grained' as to impose a binary distinction between complete overlap and complete non-overlap of momentum- and position-space wave packet envelopes; then Eq. (\ref{fullNorm}) goes over to  
\begin{eqnarray}
\left\langle \Phi_N | \Phi_N\right\rangle \rightarrow 1 
&+&  \sum_{{\cal P}\ne 1} \delta_{\cal P} \prod_{a=1}^N  \delta_{\alpha_{{\cal P}a} \alpha_a}
(2\pi)^3 \delta^3({\bf p}_{{\cal P}a} - {\bf p}_a) \nonumber \\
& &\times \,\delta^3({\bf x}_{{\cal P}a,0} - {\bf x}_{a,0}). \label{classicalNorm}
\end{eqnarray}
Here the normalizations associated with the momentum and position $\delta$ functions mirror Eqs. (\ref{momentumNormalization}) and (\ref{positionNormalization}) respectively; in particular the $\delta$ functions of zero argument are to be interpreted as $(2\pi)^3 \delta^3({\bf p}_{{\cal P}a} - {\bf p}_a)|_{{\bf p}_{{\cal P}a} = {\bf p}_a} = V$ and $\delta^3({\bf x}_{{\cal P}a,0} - {\bf x}_{a,0})|_{{\bf x}_{{\cal P}a,0} = {\bf x}_{a,0}} = V^{-1}$, where $V$ is the (effectively infinite) quantization volume. Hence the exclusion principle is promoted from strict applicability to global momentum eigenstates to `for all practical purposes' joint applicability to the centroids of momentum- and position-space wave packets \footnote{While this work is ultimately directed towards neutrino transport, it is evident that a physical picture like that discussed here must also underpin the approximation of `local thermodynamic equilibrium,' in which Fermi statistics is taken to apply to volume elements that are `microscopically large' but `macroscopically small,' and thermodynamic quantities like density, temperature, chemical potentials, and so on are taken to be continuous functions of space and time.}. (Harking back to the paragraph before last, the factor $\delta_{\alpha_{{\cal P}a} \alpha_a}$ pertains to the special initial condition in which all neutrinos have definite flavor at $t=0$; more generally, this Kronecker $\delta$ would be replaced by $\sum_{i_a} c_{\alpha_{{\cal P}a} i_a}^* c_{\alpha_a i_a}$.)

With this sort of multiparticle state in mind the expectation values in Eqs. (\ref{relativisticNeutrinoExpectationValue0}) and (\ref{relativisticAntineutrinoExpectationValue0}) can be evaluated. In particular, continuing with the example of a pure state of $N$ neutrinos given by Eq. (\ref{multineutrinoState}), the action of an annihilation operator upon $\left| \Phi_N\right\rangle$ results in a sum of $N$ terms:
\begin{equation}
a_{{\bf q}, \downarrow,i} \left| \Phi_N\right\rangle = \sum_{a=1}^N (-)^{a+1}\, \phi_{{\bf p}_a}({\bf q}) \, \mathrm{e}^{-\mathrm{i}{\bf q} \cdot {\bf x}_{a,0}}\, U_{\alpha_a i}^* \left| \Phi_{N \slashed{a}}\right\rangle, \label{annihilatedState}
\end{equation} 
where
\begin{eqnarray}
\left| \Phi_{N \slashed{a}}\right\rangle \!&\!=\!&\! \left(\prod_{b=1,b\ne a}^N \int \frac{d^3{\bf k}_b}{(2\pi)^3} \frac{\phi_{{\bf p}_b}({\bf k}_b)}{\sqrt{2 E_{{\bf k}_b}}} \mathrm{e}^{-\mathrm{i}{\bf k}_b \cdot {\bf x}_{b,0}} \sum_{i_b} U_{\alpha_b i_b}^* \right) \nonumber \\
\!&\! \!&\! \times \left| {\bf k}_1 i_1 \dots {\bf k}_{a-1} i_{a-1}, {\bf k}_{a+1} i_{a+1} \dots {\bf k}_N i_N\right\rangle \label{multineutrinoMinusState}
\end{eqnarray}
is the $N-1$ particle state resulting from the `removal' of the $a$th neutrino from $\left| \Phi_N\right\rangle$.
The sum in Eq. (\ref{annihilatedState}) arises because of the anticommutation rule obeyed by neutrino creation and annihilation operators, together with the fact (mentioned above) that the multiparticle momentum eigenstates $\left| {\bf k}_1 i_1 \dots {\bf k}_N i_N\right\rangle$ are constructed by acting upon the vacuum with antisymmetrized products of neutrino creation operators $a^\dagger_{{\bf k},\downarrow,i}$. (In particular, the relation
\begin{eqnarray}
\lefteqn{a_{{\bf q}, \downarrow,i} \left| {\bf k}_1 i_1 \dots {\bf k}_N i_N\right\rangle} \nonumber \\
&=&  \sum_{a=1}^N (-)^{a+1}\, (2\pi)^3 \, \sqrt{2E_{\bf q}} \; \delta^3({\bf q} - {\bf k}_a) \;\delta_{i i_a} \nonumber \\
& & \times \left| {\bf k}_1 i_1 \dots {\bf k}_{a-1} i_{a-1}, {\bf k}_{a+1} i_{a+1} \dots {\bf k}_N i_N\right\rangle
\end{eqnarray}
has been employed in obtaining Eqs. (\ref{annihilatedState}) and (\ref{multineutrinoMinusState}).) The expectation value in Eq. (\ref{relativisticNeutrinoExpectationValue0}), which follows from Eq. (\ref{annihilatedState}), is
\begin{eqnarray}
\langle a_{{\bf u}, \downarrow,j}^\dagger a_{{\bf q}, \downarrow,i}\rangle &=& \sum_{b=1}^N \sum_{a=1}^N (-)^{b+1} (-)^{a+1} \phi_{{\bf p}_b}({\bf u}) \phi_{{\bf p}_a}({\bf q}) \nonumber \\
& & \times \,\mathrm{e}^{\mathrm{i}{\bf u} \cdot {\bf x}_{b,0}}\,  \mathrm{e}^{-\mathrm{i}{\bf q} \cdot {\bf x}_{a,0}}\, U_{\alpha_b j}\, U_{\alpha_a i}^* \nonumber \\
& & \times \, \left\langle \Phi_{N\slashed{b}} | \Phi_{N\slashed{a}} \right\rangle / \left\langle \Phi_N | \Phi_N\right\rangle. \label{expectationValue}
\end{eqnarray}
Evaluate the norms $\left\langle \Phi_{N\slashed{b}} | \Phi_{N\slashed{a}} \right\rangle$ and $\left\langle \Phi_N | \Phi_N\right\rangle$ in the coarse-grained `all-or-nothing' view of wave packet overlap that led to Eq. (\ref{classicalNorm}), and let all the neutrino flavors and wave packet momentum centroids and positions be non-overlapping as required for $\left\langle \Phi_N | \Phi_N\right\rangle \rightarrow 1$ instead of $\left\langle \Phi_N | \Phi_N\right\rangle \rightarrow 0$. In this same approximation $\left\langle \Phi_{N \slashed{b}}| \Phi_{N \slashed{a}}\right\rangle \rightarrow \delta_{ab}$. Hence Eq. (\ref{expectationValue}) becomes
\begin{eqnarray}
\langle a_{{\bf u}, \downarrow,j}^\dagger a_{{\bf q}, \downarrow,i}\rangle &\rightarrow& \sum_{a=1}^N  \phi_{{\bf p}_a}({\bf u}) \phi_{{\bf p}_a}({\bf q}) \nonumber \\
& & \times \, \mathrm{e}^{\mathrm{i}{\bf u} \cdot {\bf x}_{a,0}}\,  \mathrm{e}^{-\mathrm{i}{\bf q} \cdot {\bf x}_{a,0}} \, U_{\alpha_a j}\, U_{\alpha_a i}^*  \label{expectationValueClassical}
\end{eqnarray}
when the norms are evaluated in the coarse-grained position/momentum picture, with all of the momentum and position wave packet centroids ${\bf p}$ and ${\bf x}_0$ being different by at least a wave packet width or so.

Before applying the coarse-grained  `all-or-nothing' view of wave packet overlap to the remaining wave packets in Eq. (\ref{expectationValueClassical}) it is appropriate to consider the evolution of these wave packets that follows from putting this expression for $\langle a_{{\bf u}, \downarrow,j}^\dagger a_{{\bf q}, \downarrow,i}\rangle$ back into the density function's mixed representation of Eq. (\ref{neutrinoDensityFunction}), via Eq. (\ref{relativisticNeutrinoExpectationValue0}). In this connection it is convenient to `open up' the $\delta$ function in Eq. (\ref{neutrinoDensityFunction}):
\begin{equation}
(2\pi)^4\,\delta^4\left(p - \frac{q_i+u_j}{2}\right) = \int d^4\Xi\; \mathrm{e}^{\mathrm{i}\left(p - \frac{q_i+u_j}{2}\right)\cdot\Xi}. \label{deltaFunction}
\end{equation}
According to the usual wave packet technology, the integral over ${\bf q}$ in each of the $N$ terms in Eq. (\ref{relativisticNeutrinoExpectationValue0}) yields a moving position-space wave packet:
\begin{eqnarray}
\lefteqn{\int \frac{d^3{\bf q}}{(2\pi)^3}\; \phi_{{\bf p}_a}({\bf q})\;
\mathrm{e}^{-\mathrm{i} q_{i} \cdot \left(x+\frac{\Xi}{2}\right)} \; \mathrm{e}^{-\mathrm{i}{\bf q}\cdot {\bf x}_{a,0}} \; \xi_{\bf q}^\downarrow } 
\nonumber \\ 
&\approx& \psi_{{\bf x}_a(t+\Xi^0/2)-{\bf \Xi}/2}({x})\; \mathrm{e}^{-\mathrm{i} p_{a,i} \cdot \left(x+\frac{\Xi}{2}\right)} \; \mathrm{e}^{-\mathrm{i}{\bf p}_a\cdot {\bf x}_{a,0}} \; \xi_{{\bf p}_a}^\downarrow. \label{qIntegral}
\end{eqnarray}
Here $(p_{a,i}^\mu)=(E_{{\bf p}_a,i},{\bf p}_a)$, and
\begin{equation}
{\bf x}_a(t) = {\bf x}_{a,0} + {\bf v}_{a} t, \label{classicalTrajectory}
\end{equation}
where ${\bf v}_{a} \equiv {\bf p}_{a}/E_{{\bf p}_a,i}\approx {\bf p}_{a}/|{\bf p}_{a}|$ is the wave packet's group velocity, approximated to ${\cal O}(m_\nu^2/ E_\nu)$. The wave packet centroid does not follow this classical trajectory for generic values of $\Xi$ (all of which are probed, thanks to the integral over $\Xi$ in Eq. (\ref{deltaFunction})); instead, the notation $\psi_{{\bf x}_a(t+\Xi^0/2)+{\bf \Xi}/2}({x})$ for the wave packet envelope is meant to convey the fact that the centroid follows the non-classical trajectory 
\begin{equation}
{\bf x}_a\left(t+\frac{\Xi^0}{2}\right)- \frac{{\bf \Xi}}{2}= {\bf x}_{a,0} - \frac{{\bf \Xi}}{2}+ {\bf v}_{a} \left(t + \frac{\Xi^0}{2}\right).
\end{equation}
Similarly, the integral over ${\bf u}$ in each of the $N$ terms in Eq. (\ref{relativisticNeutrinoExpectationValue0}) yields a factor
\begin{eqnarray}
\lefteqn{\int \frac{d^3{\bf u}}{(2\pi)^3}\; \phi_{{\bf p}_a}({\bf u})\;
\mathrm{e}^{\mathrm{i} u_{j} \cdot \left(x-\frac{\Xi}{2}\right)} \; \mathrm{e}^{\mathrm{i}{\bf u}\cdot {\bf x}_{a,0}} \; {\xi_{\bf u}^\downarrow}^\dagger } 
\nonumber \\ 
&\approx& \psi^*_{{\bf x}_a(t-\Xi^0/2)+{\bf \Xi}/2}({x})\; \mathrm{e}^{\mathrm{i} p_{a,j} \cdot \left(x-\frac{\Xi}{2}\right)} \; \mathrm{e}^{\mathrm{i}{\bf p}_a\cdot {\bf x}_{a,0}} \; {\xi_{{\bf p}_a}^\downarrow}^\dagger, \label{uIntegral}
\end{eqnarray}
which has three differences from Eq. (\ref{qIntegral}): complex conjugation, the replacement $\Xi\rightarrow -\Xi$, and mass index $j$ instead of $i$.

Now that the evolution of the wave packets has been considered the full expression for the mixed representation of the density function can be evaluated. 

The first step is to apply the classicality conditions of Eq. (\ref{classicalityConditions}), which can be related to assumptions of `slow change' and `weak inhomogeneity.' Use of Eqs. (\ref{relativisticNeutrinoExpectationValue0}),  (\ref{expectationValueClassical}), (\ref{deltaFunction}), (\ref{qIntegral}), and (\ref{uIntegral}) in Eq. (\ref{neutrinoDensityFunction}) yields 
\begin{eqnarray}
-\mathrm{i}\, G^{LR}_{ij}(x,p) &\approx& \sum_{a=1}^N \int d^4\Xi \; D_a(x)\, \mathrm{e}^{\mathrm{i}\left(p - \frac{p_{a,i}+p_{a,j}}{2}\right)\cdot\Xi}  \nonumber \\
& &\times \; \mathrm{e}^{-\mathrm{i}(E_{{\bf p}_a,i}-E_{{\bf p}_a,j})t} \,U_{\alpha_a j}\, U_{\alpha_a i}^* \nonumber \\
& &\times \;  \xi_{{\bf p}_a}^\downarrow {\xi_{{\bf p}_a}^\downarrow}^\dagger, \label{neutrinoDensityFunction1}
\end{eqnarray}
where
\begin{equation}
D_a(x) \equiv  \psi^*_{{\bf x}_a(t-\Xi^0/2)+{\bf \Xi}/2}({x})\; \psi_{{\bf x}_a(t+\Xi^0/2)-{\bf \Xi}/2}({x}).
\end{equation}
Because of the wave packets' localization, $D_a(x)$ peaks at $\Xi = 0$, for which the centroids of both wave packet envelopes follow the classical trajectory of Eq. (\ref{classicalTrajectory}). Expanding about $\Xi=0$,
\begin{equation}
D_a(x) = \left| \psi_{{\bf x}_a(t)}({x}) \right|^2 + \Xi\cdot \left[\left. \frac{\partial}{\partial x} D_a(x) \right|_{\Xi=0}\right] + \dots \label{spacetimeWavepacketOverlap}
\end{equation}
Note that the first correction term, in combination with the first exponential in Eq. (\ref{neutrinoDensityFunction1}), can be expressed
\begin{eqnarray}
\lefteqn{\Xi\cdot \left[\left. \frac{\partial}{\partial x} D_a(x) \right|_{\Xi=0}\right] \mathrm{e}^{\mathrm{i}\left(p - \frac{p_{a,i}+p_{a,j}}{2}\right)\cdot\Xi} } \nonumber \\
\!&\!=\!&\! -\mathrm{i} \frac{\partial}{\partial p}\cdot \left\{\left[\left. \frac{\partial}{\partial x} D_a(x) \right|_{\Xi=0}\right] \mathrm{e}^{\mathrm{i}\left(p - \frac{p_{a,i}+p_{a,j}}{2}\right)\cdot\Xi} \right\}. \label{correction}
\end{eqnarray}
In a collection of particles satisfying the classicality conditions of Eq. (\ref{classicalityConditions}), it can be expected on dimensional grounds that this and higher-order corrections can be neglected in the sum over a large number of particles $N$ in Eq. (\ref{neutrinoDensityFunction1}), thanks to the combination of derivatives in Eq. (\ref{correction}).

A few more simple steps bring the position and momentum dependence of Eq. (\ref{neutrinoDensityFunction1}) into fully classical form. Keeping only the first term in Eq. (\ref{spacetimeWavepacketOverlap}), the integral over $\Xi$ in Eq. (\ref{neutrinoDensityFunction1}) yields a four-momentum $\delta$ function (and integration over $p^0$ reduces this to a three-momentum $\delta$ function). Moreover, in the coarse-grained view of `all-or-nothing' wave packet overlap, this first term of $D_a(x)$ approaches a sharp restriction to the classical trajectory of Eq. (\ref{classicalTrajectory}):
\begin{equation}
D_a(x) \approx \left| \psi_{{\bf x}_a(t)}({x}) \right|^2 \rightarrow \delta^3({\bf x} - {\bf x}_a(t)).
\end{equation}
Finally, because the two spinors in the outer product $\xi_{{\bf p}_a}^\downarrow {\xi_{{\bf p}_a}^\downarrow}^\dagger$ pertain to the same momentum, Eq. (\ref{equalMomentumSpinors}) applies; hence taking the trace over this $2\times 2$ block in spinor index space results in a factor of unity. All together, the neutrino distribution matrix obtained from Eq. (\ref{neutrinoDensityFunction1}) is
\begin{eqnarray}
\rho_{ij}(t,{\bf x},{\bf p}) &=&\int \frac{dp^0}{2\pi} {\rm Tr}\left[-\mathrm{i} \,G^{LR}_{ij}(t,{\bf x},p)\right]  \label{classicalNeutrinoDistribution0} \\
&\rightarrow& \sum_{a=1}^N (2\pi)^3\, \delta^3({\bf x} - {\bf x}_a(t))\; \delta^3({\bf p} - {\bf p}_a) \nonumber \\
& & \times \; \mathrm{e}^{-\mathrm{i}(E_{{\bf p}_a,i}-E_{{\bf p}_a,j})t} \,U_{\alpha_a j}\, U_{\alpha_a i}^*. \label{classicalNeutrinoDistribution}
\end{eqnarray}
A similar argument applies to antineutrinos, with the result 
\begin{eqnarray}
\bar\rho_{ij}(t,{\bf x},{\bf p}) &=&\int \frac{dp^0}{2\pi} {\rm Tr}\left[-\mathrm{i} \,\bar G^{LR}_{ij}(t,{\bf x},p)\right]  \label{classicalAntineutrinoDistribution0} \\
&\rightarrow& \sum_{a=1}^N (2\pi)^3\, \delta^3({\bf x} - {\bf x}_a(t))\; \delta^3({\bf p} - {\bf p}_a) \nonumber \\
& & \times \; \mathrm{e}^{\mathrm{i}(E_{{\bf p}_a,i}-E_{{\bf p}_a,j})t} \,U_{\alpha_a j}\, U_{\alpha_a i}^* \label{classicalAntineutrinoDistribution}
\end{eqnarray}
for the antineutrino distribution matrix. Here ${\bf p}$ takes on the values of momentum-space wave packet centroids (which ultimately correspond to the momenta of classical particles), in accordance with the role denoted by ${\bf p}$ in Sec. \ref{sec:introduction}.

The position and momentum dependence of Eqs. (\ref{classicalNeutrinoDistribution}) and (\ref{classicalAntineutrinoDistribution}) is precisely that of a collection of free classical particles, and the associated species-space structure is just as expected. These expressions were derived in terms of mass fields, because these represent the `physical particles' for which creation and annihilation operators obeying appropriate anticommutation relations exist for arbitrary momentum. But once an average is taken for a system containing only relativistic neutrinos, the notion of a `flavor basis' for the distribution matrices becomes workable \footnote{Some authors introduce creation and annihilation operators as quanta of flavor fields, but because these do not have appropriate anticommutation relations for arbitrary momenta it is conceptually more correct to wait to introduce a `flavor basis' for the distribution matrix until an average is taken for a system containing only relativistic neutrinos. It is not that flavor fields are not well-defined; it is simply that these cannot be expanded in terms of creation and annihilation operators in momentum space, and therefore no well-defined flavor number operator (which would contain a sum over all momenta) can be defined in momentum space either.}. From Eq. (\ref{densityFunction}) it is apparent that density function $\Gamma_{\alpha\beta}^{\ell m}(y,z)$ constructed from flavor fields $\nu_\alpha(y) = U_{\alpha i} \, \nu_i(y)$ is related to that constructed from mass fields by
\begin{equation}
\Gamma_{\alpha\beta}^{\ell m}(y,z) = \sum_{i,j} U_{\alpha i} \, \Gamma_{ij}^{\ell m}(y,z) \, U_{\beta j}^*. 
\end{equation}
This propagates down to a `flavor basis' representation of the distribution matrices:
\begin{eqnarray}
\rho_{\alpha\beta}(t,{\bf x},{\bf p}) &=& \sum_{i,j} U_{\alpha i} \, \rho_{ij}(t,{\bf x},{\bf p}) \, U_{\beta j}^*, \label{neutrinoFlavorBasis} \\
\bar\rho_{\alpha\beta}(t,{\bf x},{\bf p}) &=& \sum_{i,j} U_{\alpha i} \, \bar\rho_{ij}(t,{\bf x},{\bf p}) \, U_{\beta j}^*.  \label{antineutrinoFlavorBasis}
\end{eqnarray}
In particular the diagonal elements can be expressed
\begin{eqnarray}
\rho_{\beta\beta}(t,{\bf x},{\bf p}) &\rightarrow& \sum_{a=1}^N \sum_i (2\pi)^3\, \delta^3({\bf x} - {\bf x}_a(t))\; \delta^3({\bf p} - {\bf p}_a) \nonumber \\
& & \times \; \left| U_{\beta i}\, \exp\left(-\mathrm{i}\frac{m_i^2 \,t}{2|{\bf p}_a|} \right) \,U_{\alpha_a i}^* \right|^2, \\ \label{neutrinoflavorBasisDiagonal}
\bar\rho_{\beta\beta}(t,{\bf x},{\bf p}) &\rightarrow& \sum_{a=1}^N \sum_i (2\pi)^3\, \delta^3({\bf x} - {\bf x}_a(t))\; \delta^3({\bf p} - {\bf p}_a) \nonumber \\
& & \times \; \left| U_{\beta i}^*\, \exp\left(-\mathrm{i}\frac{m_i^2 \,t}{2|{\bf p}_a|} \right) \, U_{\alpha_a i}\right|^2. \label{antineutrinoflavorBasisDiagonal}
\end{eqnarray}
These expressions describe collections of neutrinos and antineutrinos that begin in flavors $\alpha_a$ at $t=0$ and follow classical spacetime trajectories, with the expectation values of their flavors along those trajectories varying according to familiar vacuum oscillation probabilities \cite{Yao2006Review-of-Parti}.

To conclude this subsubsection, it is appropriate to remark on the illustrative character of this derivation of the classical limit of the position/momentum dependence. A single pure state of neutrinos has been singled out for detailed discussion, and the average denoted by angled brackets in Eq. (\ref{relativisticNeutrinoExpectationValue0}) has been interpreted as an expectation value with respect to this pure neutrino state. Hence the derivation of Eqs. (\ref{classicalNeutrinoDistribution}) and (\ref{classicalAntineutrinoDistribution}) constitutes a demonstration of how a single `microstate' of uncorrelated single-particle quantum wave packets corresponds to a single `microstate' of definite classical trajectories (while retaining the quantum mechanical phenomena of flavor mixing and Fermi statistics). This adequately illustrates the issues involved in a passage to the classical limit. But it should also be noted that in order for $\rho(t,{\bf x},{\bf p})$ and $\bar\rho(t,{\bf x},{\bf p})$ to actually be single-particle distribution matrices for the `macrostate' of the gas, the average represented by angle brackets must be promoted to an average over a {\em mixed} state of neutrinos or antineutrinos respectively, together with an ensemble average over an appropriate statistical distribution of these mixed states. (The mixed states of neutrinos or antineutrinos, represented by density matrices spanning all possible numbers of neutrinos or antineutrinos with all possible independent single-particle wave packets, would be obtained by integration of the pure states of the entire system over the degrees of freedom of all particle types other than neutrinos or antineutrinos, in accordance with the standard meaning of mixed states and density matrices \cite{Landau1977Quantum-Mechani}.) Then, in an appropriate limit along the lines presented here, the position/momentum dependence of the single-particle distribution matrices would correspond to the position/momentum dependence of single-particle distribution functions for a statistical ensemble of classical particles.
Finally, as a last reminder of another illustrative feature, the factor $U_{\alpha_a j}\, U_{\alpha_a i}^*$ in Eqs. (\ref{classicalNeutrinoDistribution}) and (\ref{classicalAntineutrinoDistribution}) corresponds to the special case in which the neutrinos or antineutrinos respectively have definite flavors at $t=0$. In more general cases one would have $c_{\alpha_a j}^*\, c_{\alpha_a i}$, where the meaning of the amplitude $c_{\alpha_a i}$ was explained in the fifth paragraph of this subsubsection.

\subsection{Liouville equations}

The Liouville equations for the neutrino and antineutrino distribution matrices of Eqs. (\ref{classicalNeutrinoDistribution0}), (\ref{classicalAntineutrinoDistribution0}) (`mass basis') and (\ref{neutrinoFlavorBasis}), (\ref{antineutrinoFlavorBasis}) (`flavor basis') follow immediately from Eqs. (\ref{neutrinoDensityEquation}) and (\ref{antineutrinoDensityEquation}). In matrix form, with species indices suppressed, the Liouville equations are those obtained at the end of Sec. \ref{sec:simple}:
\begin{eqnarray}
p^\mu \frac{\partial}{\partial x^\mu} \rho(t,{\bf x},{\bf p}) + \frac{\mathrm{i}}{2} \left[  \Delta, \rho(t,{\bf x},{\bf p}) \right] &=& 0, \label{neutrinoLiouville3}\\
p^\mu \frac{\partial}{\partial x^\mu} \bar\rho(t,{\bf x},{\bf p}) - \frac{\mathrm{i}}{2} \left[ \Delta, \bar\rho(t,{\bf x},{\bf p}) \right] &=& 0. \label{antineutrinoLiouville3}
\end{eqnarray}
As mentioned immediately following Eq. (\ref{classicalTrajectory}), the trajectories $\mathbf{x}_a(t)$ appearing in Eqs. (\ref{classicalNeutrinoDistribution}) and (\ref{classicalAntineutrinoDistribution}) are null to ${\cal O}(m_\nu^2/ E_\nu)$. In accordance with this---and in order for these explicit expressions for $\rho(t,{\bf x},{\bf p})$ and $\bar\rho(t,{\bf x},{\bf p})$ to satisfy Eqs. (\ref{neutrinoLiouville3}) and (\ref{antineutrinoLiouville3})---the momentum $p$ in the first term of the Liouville equations above should be approximated as $(p^\mu) = (|\mathbf{p}|, \mathbf{p})$. Only in phases in the explicit expressions for $\rho(t,{\bf x},{\bf p})$ and $\bar\rho(t,{\bf x},{\bf p})$ encountered in the previous subsection are corrections of $\mathcal{O}(m^2/ 2|{\bf p}|)$ to $E_{\bf p}$  retained.

\section{Conclusion}
\label{sec:conclusion}

The calculation of neutrino decoupling from dense nuclear matter requires a transport formalism capable of handling both collisions and flavor mixing; and the first steps towards such a formalism are the construction of neutrino and antineutrino `distribution matrices,' and a determination of the Liouville equations they satisfy in the noninteracting case. These initial steps have been accomplished in two new ways in this paper. Both approaches arrive at neutrino and antineutrino distribution matrices $\rho(t,{\bf x},{\bf p})$ and $\bar\rho(t,{\bf x},{\bf p})$ whose dependence on time $t$, position ${\bf x}$, and momentum ${\bf p}$ is classical. Indeed the diagonal elements of these distribution matrices are classical distribution functions, in the sense of Eq. (\ref{classicalDistribution}), for the various neutrino species. The off-diagonal elements encode information on species overlap in the neutrino ensemble.  The flat-spacetime Liouville equations satisfied by $\rho(t,{\bf x},{\bf p})$ and $\bar\rho(t,{\bf x},{\bf p})$ are given both in Eqs. (\ref{neutrinoLiouville2}), (\ref{antineutrinoLiouville2}) and (\ref{neutrinoLiouville3}), (\ref{antineutrinoLiouville3}). In addition to the usual flat-spacetime directional derivative along the phase flow $p^\mu \, \partial/\partial x^\mu$ (see also endnote \cite{endnote40}),
the Liouville operators for neutrinos and antineutrinos with flavor mixing include a commutator with a matrix containing differences of squared neutrino masses (see Eq. (\ref{massDifferenceMatrix})), with a difference in sign between the neutrino and antineutrino cases. 

In the approach of Sec. \ref{sec:simple} the neutrino positions and momenta are taken to be classical from the outset: only neutrino mass/flavor are treated quantum mechanically. Distribution matrices are constructed from the states of a covariant version of the familiar simple model of flavor mixing, and the Liouville equations follow straightforwardly from the `Schr\"odinger equations' describing the evolution of flavor along classical worldlines.
 
The second approach---presented in Sec. \ref{sec:correlation}---employs a `density function,' the mean value of paired neutrino quantum field operators (Eq. (\ref{densityFunction})); therefore the classical position/momentum dependence must be derived as a limit. The key to this is the `mixed representation' of the density function obtained by a Wigner transformation (Eq. (\ref{neutrinoWigner})). By definition the spacetime variable $x$ of the mixed representation is the average of the field operators' position variables (Eq. (\ref{backwardChange1})), and the momentum variable $P$ of the mixed representation also turns out to be (up to a sign in the case of antineutrinos) an average of the momenta appearing the field operators' plane wave expansions (note the $\delta$ functions in Eqs. (\ref{neutrinoDensityFunction}) and (\ref{antineutrinoDensityFunction})). In order to make a concrete physical connection between these density functions and distribution matrices, considerable space is devoted in Subsec. \ref{subsec:interpretation} to explicit examination of the general relationship of the mixed representation to complementary space and momentum distributions, and especially to a detailed illustration of how a suitable state of uncorrelated wave packets, satisfying basic classicality conditions of slow variation and weak inhomogeneity (Eqs. (\ref{classicalityConditions})),  corresponds to a microstate of a gas whose neutrinos follow classical trajectories while exhibiting the usual flavor oscillations. This last illustration is by far the paper's densest thicket; by comparison, almost magically simple and elegant is the emergence in Subsec. \ref{subsec:firstOrder} of the Liouville operator from a Wigner transformation of the difference of Klein-Gordon equations obeyed by the density functions.

Given the existence of the simpler approach of Sec. \ref{sec:simple}, the question arises as to why it is worth bothering with the more fundamental approach of Sec. \ref{sec:correlation}. The reason is that it is necessary to go beyond the case of noninteracting neutrino distributions that satisfy Liouville equations: neutrino interactions need to be considered, and our understanding of neutrino interactions is based on quantum field theory. It is true that one can attempt to compute neutrino effective masses and interaction rates independently and then insert them into a formalism based on the simple model of neutrino flavor mixing. But there is a history of overlooking important aspects of the problem when such an approach is taken. For instance, early works that considered contributions of neutrino-neutrino forward scattering to effective neutrino (squared) masses, such as Refs. \cite{Fuller1987Resonant-Neutri,Notzold1988Neutrino-disper}, failed to recognize the existence of off-diagonal contributions \cite{Pantaleone1992Dirac-neutrinos,McKellar1994Oscillating-neu}. Another example is an apparent failure \cite{Strack2005Generalized-Bol} to recognize that the placement of neutrino blocking factors in neutrino interaction rates is nontrivial, due to the fact that neutrino distributions are now represented by non-commuting matrices \cite{Sigl1993General-kinetic}. Such issues are automatically raised and naturally handled in a formalism that begins by treating all aspects of the problem in terms of quantum field theory from the beginning; hence, in addition to being conceptually satisfying, an approach of this kind is also theoretically safe. 

It may then further be asked why a new treatment might be desirable if in fact interactions have already been responsibly addressed in the literature \cite{Sigl1993General-kinetic}. While the handling of general classes of interactions is outlined in Ref. \cite{Sigl1993General-kinetic}, the particular interactions relevant to neutrino decoupling are not spelled out with the degree of explicit specificity needed by those developing large-scale simulations involving neutrino transport. In addition to the present work's goal of more thorough understandings of the spatial derivative term in the Liouville equation and the nature of neutrino distribution matrices, it also serves as a first step towards the handling of neutrino interactions with a diagrammatic approach based on a nonequilibrium Green's function \cite{Lifshitz1981Physical-Kineti} (see also Ref. \cite{Yamada2000Boltzmann-equat}). In addition to a Green's function, this approach---which is sometimes called `Keldysh theory'---involves the density function considered in this paper and two other types of field operator pairings. Because it is a diagrammatic approach while that of Ref. \cite{Sigl1993General-kinetic} is purely algebraic, it can be expected to simplify the task of explicitly working out the full panoply of interactions needed by those wishing to incorporate neutrino flavor mixing into neutrino transport computations. This will be pursued in a separate work.


\appendix*
\section{}

The approach in Sec. \ref{sec:correlation} to neutrino distribution matrices and the Liouville equations they obey is rooted in quantum field theory. Quantum `density functions' constructed from neutrino fields simplify considerably in the relativistic limit of practical interest.

Neutrino interactions respect lepton number in the Standard Model, but with good evidence from flavor mixing observations and experiments that neutrinos have mass, it is not clear that neutrinos actually carry lepton number \cite{Yao2006Review-of-Parti}. Standard Model neutrino interactions are of the $V-A$ form and therefore involve only left-handed neutrino fields. For massless neutrinos this implies the existence of only two neutrino states for each flavor: negative-helicity neutrinos and positive-helicity antineutrinos. In the relativistic limit of extensions that give the neutrinos mass, these two states could be either left-handed `neutrino' and right-handed `antineutrino' states associated with a Dirac field, or the left- and right-handed states of a self-conjugate Majorana field carrying no net lepton number. Differences between these two possibilities occur only at ${\cal O}(m_\nu^2/ E_\nu^2)$, where $m_\nu$ and $E_\nu$ are characteristic neutrino mass and energy scales. Systems in which effects at this scale are relevant are not considered in this paper; instead, only terms of ${\cal O}(m_\nu^2/ E_\nu)$ are kept, which capture the flavor mixing physics relevant to the neutrino decoupling problem. For definiteness the language and formalism of Dirac fields are used here---that is, `neutrinos' and `antineutrinos' and associated operators $a_{{\bf q}, r}, b_{{\bf q}, r}$ and so forth are spoken of---but the results to ${\cal O}(m_\nu^2/ E_\nu)$ are the same as if Majorana fields were employed. 

The operator $\psi^{\ell}(y)$ representing a noninteracting spin-1/2 field possessing one or more quantum numbers distinguishing particles from antiparticles---that is, a Dirac field---is
\begin{equation}
\psi^{\ell}(y) = A^\ell(y) + B^\ell(y),
\end{equation}
where $A^\ell(y)$ and $B^\ell(y)$ are functions of spacetime position $y$ and constitute `positive-' and `negative-frequency' parts respectively:
\begin{eqnarray}
A^{\ell}(y) \!\!\!&=&\!\!\! \int \frac{d^3{\bf q}}{(2\pi)^3} \frac{1}{\sqrt{2 E_{\bf q}}} 
	\sum_r u^\ell({\bf q}, r)\, \mathrm{e}^{-\mathrm{i} q\cdot y}\,a_{{\bf q}, r}, \label{positiveFrequency}\\
B^\ell(y) \!\!\!&=&\!\!\! \int \frac{d^3{\bf q}}{(2\pi)^3} \frac{1}{\sqrt{2 E_{\bf q}}} 
	\sum_r v^\ell({\bf q}, r) \, \mathrm{e}^{\mathrm{i} q\cdot y} \,b_{{\bf q}, r}^\dagger. \label{negativeFrequency}
\end{eqnarray}
Here $\ell$ is a spinor index and $r$ labels spin states. The momentum 4-vector $q$ has components $(q^\mu) =  (E_{\bf q}, {\bf q})$, where $E_{\bf q} = \sqrt{|{\bf q}|^2 + m^2}$ is the on-shell energy of a particle of mass $m$. The positive-frequency term contains particle annihilation operators $a_{{\bf q}, r}$ and momentum-space Dirac spinors $u^\ell({\bf q}, r)$, and the negative-frequency term contains antiparticle creation operators $b_{{\bf q}, r}^\dagger$ and momentum-space Dirac spinors $v^\ell({\bf q}, r)$. The free field satisfies the Dirac equation
\begin{equation}
\left(\mathrm{i}\gamma^\mu\frac{\partial}{\partial y^\mu} - m\right)\psi(y).
\end{equation}
The Dirac spinor indices on $\gamma^\mu$ and $\psi(y)$ have been suppressed. Here the Dirac matrices $\gamma^\mu$ satisfy the anticommutation relations $\{\gamma^\mu,\gamma^\nu\}=2\eta^{\mu\nu}$, where $\eta^{\mu\nu}$ is the Lorentz metric. The conventions of Ref. \cite{Peskin1995An-Introduction} are followed for units ($\hbar = c= 1$); metric signature ($+---$); creation/annihilation operator anticommutation relations [$\{a_{{\bf q},r},a_{{\bf u},s}^\dagger\} = \{b_{{\bf q},r},b_{{\bf u},s}^\dagger\} = (2\pi)^3 \delta^3({\bf q}-{\bf u})\delta_{rs}$, with all other anticommutators vanishing]; single-particle states ($| {\bf q}, r\rangle \equiv \sqrt{2E_{\bf q}}\,a_{{\bf q},r}^\dagger |0\rangle$) and their normalization [$\langle {\bf q}, r | {\bf u}, s\rangle = 2E_{\bf q} (2\pi)^3 \delta^3({\bf q}-{\bf u})\delta_{rs}$]; and Dirac matrices, which have
the $2\times 2$ block form
\begin{equation}
(\gamma^\mu) = \pmatrix{0 &  \sigma^\mu \cr \bar\sigma^\mu & 0}.
\end{equation}
Here $(\sigma^\mu) = (1,\bm{\sigma})$ and $(\bar\sigma^\mu) = (1,-\bm{\sigma})$, where $\bm{\sigma}$ are the standard $2\times 2$ Pauli matrices. Moreover the matrix
\begin{equation}
(\gamma^5) = \pmatrix{-1 & 0 \cr 0 & 1}
\end{equation}
appears in the left and right projection operators
\begin{eqnarray}
P_L &=& \frac{1}{2}(1-\gamma^5), \label{projectionLeft}\\
P_R &=& \frac{1}{2}(1+\gamma^5), \label{projectionRight}
\end{eqnarray}
which select the upper two and lower two components of Dirac spinors respectively.

The simplifications incident to the relativistic limit are manifest in explicit expressions for the neutrino and antineutrino momentum-space spinors. These can be written in block form as
\begin{eqnarray}
u({\bf q}, r) &=& \pmatrix{\sqrt{q\cdot\sigma}\, \xi_{\bf q}^r \cr \sqrt{q\cdot\bar\sigma}\;\xi_{\bf q}^r}, \\
v({\bf q}, r) &=& \pmatrix{\sqrt{q\cdot\sigma}\, \eta_{\bf q}^r \cr -\sqrt{q\cdot\bar\sigma}\;\eta_{\bf q}^r}.
\end{eqnarray}
It is convenient to define the two-component spinors $\xi_{\bf q}^r$ and $\eta_{\bf q}^r$ in terms of eigenspinors $\chi_{\bf q}^\pm$ that satisfy the relations
\begin{equation}
(\hat{\bf q}\cdot\bm{\sigma}) \, \chi_{\bf q}^\pm = \pm\chi_{\bf p}^\pm.\label{eigenspinors}
\end{equation}
The Dirac spinors are associated with left-handed (negative-helicity $\downarrow$) and right-handed (positive-helicity $\uparrow$) spin states through the following assignments:
\begin{eqnarray}
\xi_{\bf q}^\uparrow &=& \chi_{\bf q}^+, \\
\xi_{\bf q}^\downarrow &=& \chi_{\bf q}^-, \\
\eta_{\bf q}^\uparrow &=& \chi_{\bf q}^-, \\
\eta_{\bf q}^\downarrow &=& -\chi_{\bf q}^+.
\end{eqnarray}
In the relativistic limit the neutrino momentum spinors become
\begin{eqnarray}
u({\bf q}, \uparrow) &\rightarrow& \pmatrix{0 \cr \sqrt{2E_{\bf q}}\;\xi_{\bf p}^\uparrow}, \\
u({\bf q}, \downarrow) &\rightarrow& \pmatrix{\sqrt{2E_{\bf q}}\;\xi_{\bf p}^\downarrow \cr 0}, \label{relativisticNeutrinoSpinor}
\end{eqnarray}
while the antineutrino momentum spinors become
\begin{eqnarray}
v({\bf q}, \uparrow) &\rightarrow& \pmatrix{\sqrt{2E_{\bf q}}\;\eta_{\bf q}^\uparrow \cr 0}, \label{relativisticAntineutrinoSpinor} \\
v({\bf q}, \downarrow) &\rightarrow& \pmatrix{0 \cr \sqrt{2E_{\bf q}}\;\eta_{\bf q}^\downarrow}.
\end{eqnarray}
As mentioned above, the fact that neutrino interactions involve only left-handed neutrino fields $P_L\,\nu(y)$ implies that to ${\cal O}(m_\nu^2/ E_\nu)$ only negative-helicity neutrinos and positive-helicity antineutrinos are produced.

Consider the density function defined in Eq. (\ref{densityFunction}); see also Eq. (\ref{densityFunction2}).
Applying Eqs. (\ref{positiveFrequency}) and (\ref{negativeFrequency}) to neutrino fields with mass indices $i,j$, it can be expressed 
\begin{eqnarray}
{\rm i}\, \Gamma^{\ell m}_{ij}(y,z) &=& - \int  \frac{d^3{\bf q}}{(2\pi)^3} \frac{d^3{\bf u}}{(2\pi)^3}  \left[\mathrm{e}^{{\rm i} (u_j\cdot z - q_i\cdot y)} \mathrm{N}_{ij}^{\ell m}({\bf q},{\bf u}) \right.\nonumber \\
& & \left. \ \ \ \ \ \  - \  \mathrm{e}^{{\rm i} (q_i\cdot y - u_j\cdot z)} \bar \mathrm{N}_{ij}^{\ell m}({\bf q},{\bf u})\right].
\end{eqnarray}
where
\begin{eqnarray}
\mathrm{N}_{ij}^{\ell m}({\bf q},{\bf u}) &=& \frac{1}{\sqrt{2 E_{\bf q}}} \frac{1}{\sqrt{2 E_{\bf u}}}  \sum_{r,s} u^\ell({\bf q}, r, i) \bar u^m({\bf u}, s, j) \nonumber \\
& &\times  \langle a_{{\bf u}, s,j}^\dagger a_{{\bf q}, r,i}\rangle, \label{neutrinoExpectationValue}\\
\bar \mathrm{N}_{ij}^{\ell m}({\bf q},{\bf u}) &=& \frac{1}{\sqrt{2 E_{\bf q}}} \frac{1}{\sqrt{2 E_{\bf u}}}  \sum_{r,s} v^\ell({\bf q}, r, i) \bar v^m({\bf u}, s, j) \nonumber \\
& &\times  \langle b_{{\bf q}, r, i}^\dagger b_{{\bf u}, s, j}\rangle. \label{antineutrinoExpectationValue}
\end{eqnarray}
In these expressions the four-momentum $(u^\mu)=(u^0,{\bf u})$ should not be confused with the momentum-space Dirac spinors $u^\ell({\bf q}, r, i)$. The components of the Pauli conjugate spinors are $\bar u^m = \sum_n (u^*)^n(\gamma^0)^{n m}$ and similarly for $\bar v^m$.

To ${\cal O}(m_\nu^2/ E_\nu)$ only a single $2\times 2$ block of the $4\times4$ spinor-space structure remains nonzero, and only one of the two spin states contributes to the expectation value. In particular the nonzero $2\times 2$ blocks $\mathrm{N}_{ij}^{LR}({\bf q},{\bf u})$ and $ \bar \mathrm{N}_{ij}^{LR}({\bf q},{\bf u})$ are those that would be projected out if  $(\mathrm{N}_{ij}^{\ell m}({\bf q},{\bf u}) )$ and $(\bar \mathrm{N}_{ij}^{\ell m}({\bf q},{\bf u}) )$ were sandwiched between the left- and right-projection matrices $P_L$ and $P_R$.
Employing Eqs. (\ref{relativisticNeutrinoSpinor}) and (\ref{relativisticAntineutrinoSpinor}) in Eqs. (\ref{neutrinoExpectationValue}) and (\ref{antineutrinoExpectationValue}) results in the expressions
\begin{eqnarray}
\mathrm{N}_{ij}^{LR}({\bf q},{\bf u}) &\rightarrow& \xi_{\bf q}^\downarrow \, {\xi_{\bf u}^\downarrow}^\dagger \, \langle a_{{\bf u}, \downarrow,j}^\dagger a_{{\bf q}, \downarrow,i}\rangle, \label{relativisticNeutrinoExpectationValue}\\
\bar \mathrm{N}_{ij}^{LR}({\bf q},{\bf u}) &\rightarrow& \eta_{\bf q}^\uparrow \, {\eta_{\bf u}^\uparrow}^\dagger \, \langle b_{{\bf q}, \uparrow,i}^\dagger b_{{\bf u}, \uparrow,j}\rangle  \label{relativisticAntineutrinoExpectationValue}
\end{eqnarray}
for these non-zero $2\times2$ blocks. A nice form results when ${\bf u} = {\bf q}$, by virtue of the identity
\begin{equation}
\xi_{\bf q}^\downarrow \, {\xi_{\bf q}^\downarrow}^\dagger = \eta_{\bf q}^\uparrow \, {\eta_{\bf q}^\uparrow}^\dagger = \frac{q_\mu \sigma^\mu}{2E_{\bf q}}, \label{equalMomentumSpinors}
\end{equation}
with $|{\bf q}| = E_{\bf q}$ in the relativistic limit. (This identity follows from an explicit expression for $\chi_{\bf q}^-$ that satisfies Eq. (\ref{eigenspinors}):
\begin{equation}
\chi_{\bf q}^- = \pmatrix{-\mathrm{e}^{-\mathrm{i}\phi}\sin\left(\theta\over 2\right) \cr \cos\left(\theta\over 2 \right)},
\end{equation}
where the polar angle $\theta$ and azimuthal angle $\phi$ give the direction of ${\bf q}$.) Hence 
\begin{eqnarray}
\mathrm{N}_{ij}^{LR}({\bf q},{\bf q}) &=&  \frac{q_\mu \sigma^\mu}{2E_{\bf q}}\langle a_{{\bf q}, \downarrow,j}^\dagger a_{{\bf q}, \downarrow,i}\rangle, \label{qNeutrinoExpectationValue}\\
\bar \mathrm{N}_{ij}^{LR}({\bf q},{\bf q}) &=& \frac{q_\mu \sigma^\mu}{2E_{\bf q}}\langle b_{{\bf q}, \uparrow,i}^\dagger b_{{\bf q}, \uparrow,j}\rangle  \label{qAntineutrinoExpectationValue}
\end{eqnarray}
in the relativistic limit with ${\bf u} = {\bf q}$. The trace of these $2\times 2$ blocks is
\begin{eqnarray}
{\rm Tr}[\mathrm{N}_{ij}^{LR}({\bf q},{\bf q})] &=&  \langle a_{{\bf q}, \downarrow,j}^\dagger a_{{\bf q}, \downarrow,i}\rangle, \label{qNeutrinoExpectationValueTrace}\\
{\rm Tr}[\bar \mathrm{N}_{ij}^{LR}({\bf q},{\bf q})] &=& \langle b_{{\bf q}, \uparrow,i}^\dagger b_{{\bf q}, \uparrow,j}\rangle.  \label{qAntineutrinoExpectationValueTrace}
\end{eqnarray}
That the leading factor becomes unity is a consequence of the tracelessness of the Pauli matrices $\bm{\sigma}$.

\begin{acknowledgments}
George Fuller, Jun Hidaka, Huaiyu Duan, and especially Phil Amanik provided feedback on an early version of some of the calculations in this work. Oak Ridge National Laboratory is managed by UT-Battelle under contract DE-AC05-00OR22725 with the United States Department of Energy.
\end{acknowledgments}


\end{document}